\documentclass{article}
\usepackage[english]{babel}
\usepackage[utf8]{inputenc}
\usepackage{johd}
\usepackage{subcaption}
\usepackage{graphicx}

\usepackage[numbers,sort&compress]{natbib}
\setcitestyle{round}

\usepackage[table,xcdraw]{xcolor}
\usepackage{tabularx} 
\usepackage{array} 
\usepackage{multirow}
\usepackage{longtable}
\usepackage{amsmath}
\usepackage[labelfont=bf]{caption}
\usepackage[a4paper, total={4.8in, 9in}]{geometry}
\usepackage{lettrine}
\usepackage{afterpage}
\usepackage{multibib}

\newcites{supp}{Supplementary references}

\newcommand{\dropcapfont}{\fontfamily{cmr}\bfseries\fontsize{26pt}{28pt}\selectfont}
\newcommand{\dropcap}[1]{\lettrine[lines=2,lraise=0.05,findent=0.1em, nindent=0em]{{\dropcapfont{#1}}}{}}

\newcolumntype{Y}{>{\centering\arraybackslash}X}

\title{\vspace{1cm} Social Norms Offer Explanation for Inconsistent Effects of Incentives on Prosocial Behavior}

\author{Caroline Graf\thanks{Center for Philanthropic Studies, Dept. of Sociology, Vrije Universiteit Amsterdam} $^{,}$\thanks{Dept. of Donor Medicine, Sanquin Research} \hspace{0.001cm}, Eva-Maria Merz$^{*, \dagger}$, Bianca Suanet\thanks{Dept. of Sociology, Vrije Universiteit Amsterdam} \vspace{0.17cm} \\ and Pamala Wiepking\thanks{Lilly Family School of Philanthropy, Indiana University–Purdue University Indianapolis} $^{,*,}$\thanks{We are grateful to Wim de Kort and all blood donation experts from across Europe who shared their knowledge about incentives for donors in their countries with us. Moreover, we thank Mark Ottoni-Wilhelm, Mario Macis, Bianca Beersma, Joris Schröder and David Schultner for thoughtful comments on earlier versions of this draft. Correspondence concerning this working paper should be addressed to Caroline Graf (\href{mailto:c.graf@vu.nl}{c.graf@vu.nl}) or Eva-Maria Merz (\href{mailto:e.m.merz@vu.nl}{e.m.merz@vu.nl}).}}

\date{} 

\begin{document}

\maketitle

\begin{abstract} 
\noindent Incentives have surprisingly inconsistent effects when it comes to encouraging people to behave prosocially. Classical economic theory, according to which a specific behavior becomes more prevalent when it is rewarded, struggles to explain why incentives sometimes backfire. More recent theories therefore posit a reputational cost offsetting the benefits of receiving an incentive -- yet unexplained effects of incentives remain, for instance across incentive types and countries. We propose that social norms can offer an explanation for these inconsistencies. Ultimately, social norms determine the reputational costs or benefits resulting from a given behavior, and thus variation in the effect of incentives may reflect variation in norms. We implemented a formal model of prosocial behavior integrating social norms, which we empirically tested on the real-world prosocial behavior of blood donation. Blood donation is essential for many life-saving medical procedures, but also presents an ideal testing ground for our theory: Various incentive policies for blood donors exist across countries, enabling a comparative approach. Our preregistered analyses reveal that social norms can indeed account for the varying effects of financial and time incentives on individual-level blood donation behavior across 28 European countries. Incentives are associated with higher levels of prosociality when norms regarding the incentive are more positive. The results indicate that social norms play an important role in explaining the relationship between incentives and prosocial behavior. More generally, our approach highlights the potential of integrating theory from across the economic and behavioral sciences to generate novel insights, with tangible consequences for policy-making.

\end{abstract} 

\noindent\vspace{1cm} \\

\dropcap{H}umans are remarkable in the extent to which they engage in prosocial behavior -- and many communities want to encourage their members to help each other and contribute to public goods. According to classical economic theory, an obvious means to do this is by offering incentives\footnote{We define an incentive as an “extrinsically motivated reward, which can be monetary or nonmonetary and is used to motivate an action” \cite[][p.245]{chell_systematic_2018}. Extrinsic rewards are the positive \textit{external} outcomes associated with a given behavior, such as financial rewards. This is opposed to \textit{intrinsic rewards}, i.e., the positive \textit{internal} outcomes of a behavior, such as feeling fulfilled. While \textit{extrinsic} rewards range from financial benefits to status gains, we focus specifically on (a) \textit{financial} rewards (e.g., tax breaks, cash) and (b) non-financial \textit{time} rewards (i.e., time off work).} \citep{mill_principles_1882, marshall1890principles}. Empirical investigations have in fact demonstrated that incentives sometimes encourage prosocial behavior. For example, tax breaks increase charitable giving \cite{duquette_tax_2016}; lottery tickets and other rewards increase blood donations \cite{goette_blood_2020, lacetera_economic_2013}. However, incentives sometimes backfire. A range of studies has found incentives to have no effect on the amount of prosocial behavior exhibited; in some cases they have actually reduced prosociality \citep{deci_effects_1971, deci1999meta, mellstrom_crowding_2008, heyman_effort_2004, gneezy_fine_2000, gneezy_pay_2000}. For instance, high school students collecting donations took in \textit{less} when they were paid in proportion to the charitable donations they collected than when they were not paid \citep{gneezy_pay_2000}. 

Indeed, reviews both in economics \citep{gneezy_when_2011} and in the blood donation literature \citep{chell_systematic_2018} conclude that the effectiveness of incentives in encouraging prosocial behavior is strongly \textit{context-dependent}. Context effects include: (a) \textit{methodological} effects, e.g., surveys tend to find that people report not being attracted by incentives, whereas experiments and observational studies more often find positive effects of incentives on observed behavior \cite{lacetera_economic_2013}; (b) varying effects across incentive \textit{types}, e.g., lottery tickets \cite{goette_blood_2020} and time off work \cite{lacetera2013time} encouraged blood donation, but movie tickets \cite{royse1999exploring} did not; and (c) \textit{cross-country} variation in the effectiveness of incentives, e.g., recruitment rates for new blood donors were increased by coupons in the U.S. \cite{ferrari1985use}, but remained unchanged when vouchers were offered to Argentinians \citep{iajya2013effects}. What explains these inconsistencies, and under what circumstances do incentives encourage prosocial behavior?

\paragraph*{Previous theoretical explanations for effects of incentives on prosocial behavior.}

Richard Titmuss’s classic argument \cite{titmuss_gift_1971} that incentives lead to a deterioration of people’s intrinsic motivation to do good inspired theorizing about the \textit{detrimental} effects of incentives. In particular, it has been suggested that extrinsic rewards cause a decrease in intrinsic motivation, resulting in the phenomenon of “motivational crowding out” \cite{deci_effects_1971, deci1999meta}. Crowding out was said to prevent agents from indulging in altruistic feelings, as a result of introducing incentives \citep{frey_cost_1997, frey_motivation_2001}.

However, these accounts could not explain why incentives sometimes do encourage prosocial behavior, in particular when people are not observed (i.e., in private settings). Thus, another mechanism argued to play a role in determining the effect of incentives on prosocial behavior is \textit{reputational} motivation\footnote{We define reputational motivation as the motivation to act derived from the \textit{valuation of reputational outcomes} associated with a behavior, where reputational outcomes are the positive or negative evaluations of a behavior given what is considered \textit{appropriate} in a specific society.} \cite[also termed image motivation][]{ariely_doing_2009}. For instance, participants exerted \textit{more} effort to donate to a charitable organization if they received financial incentives for the amount of effort employed, \textit{if} it was \textit{not visible} to others that they were being paid \cite{ariely_doing_2009}. But when it was \textit{visible} to others, participants actually employed the same or \textit{less} effort than when they were not being paid. 

Similar findings emerged from the literature on dictator games (DG), where agents trade off financial reward for themselves (i.e., extrinsic motives) and sharing the endowment with another (intrinsic motivation to act prosocially). The classic paper by Hoffman et al. \cite{hoffman_social_1996} was one of the first to show that contributions in a DG were much lower in double-blind conditions where neither the other participants nor the experimenters were aware of participants’ choices, compared to when experimenters were able to observe the participants’ behavior. Similarly, participants behaved less prosocially when their behavior had less direct reputational consequences \cite[for instance because there were multiple dictators,][]{dana_exploiting_2007}. In sum, people are more likely to be tempted by an extrinsic reward when their reputation is not at stake.

This notion that people care about how they are perceived has also been incorporated into theory. Notably, Andreoni and Bernheim \cite{andreoni_social_2009} included reputational motivation in their model of dictator game behavior, where agents’ behavior is driven by, among other factors, a desire to be perceived as fair. Similarly, Bénabou and Tirole \cite{benabou_incentives_2006} modeled prosocial behavior as being influenced by a desire to be perceived as intrinsically and not extrinsically motivated. These models have been successful in predicting differences between public and private settings, and have offered an explanation for discrepancies between people's self-reported attitudes toward incentives and observed behavior. However, varying effects of different types of incentives and cross-country variation in the effectiveness of incentives remain unexplained. 

\paragraph*{Social norms govern reputational motivation.}

Reputation is in fact a core element of explanations for human prosociality in evolutionary biology and social psychology \cite{nowak_evolution_2005, nowak_evolution_1998, fehr2004social, morris2018common, cialdini_social_1998}. In particular, accounts from evolutionary biology posit that reputation formation enables indirect reciprocity, which is critical for the evolution of human cooperation \cite{nowak_evolution_2005}; in psychology, reputational concerns are argued to be an important driver for people to engage in individually costly (prosocial) behavior, because positive reputational outcomes (e.g., increased respect) result in long-term benefits \cite{morris2018common, cialdini_social_1998}. Crucially, reputation here is understood to be governed by \textit{social norms}\footnote{We define social norms as the informal rules that guide and/or constrain social behaviors, which determine how acceptable a given behavior is viewed by members of a society or group \cite{cialdini_social_1998}.} \citep{nowak_evolution_2005, alexander_biology_1987, sherif_psychology_1936, morris2018common, cialdini_social_1998}: "[t]he revision of an individual’s reputation depends on the social norms that establish what characterizes a good or bad action [...]" \cite[p.~242]{santos_social_2018}.

A critical feature of social norms is context dependency. In some contexts \textit{prosocial} norms may dominate, prescribing that resources should be shared with others, e.g., due to societal responsibility \cite{cialdini_social_1998} or fairness \cite{fehr_theory_1999}. In other contexts, it may be acceptable for people to behave more selfishly \cite[e.g., people do not prefer the 50-50 split in DG when earned wealth is allocated][]{cherry_hardnose_2002}. Krupka and Weber \cite{krupka2013identifying} provide striking evidence of the subtle context dependency of social norms in variations of the DG. For example, people found a 50-50 split more appropriate when initial allocations to the dictator and recipient were 50-50 than when initial allocations were 100-0, despite payoff equivalence in these two scenarios. In turn, this variation in norms was a good predictor of choices made by unrelated others in these games \cite{krupka2013identifying}. In addition to this situational context dependency, social norms also depend on \textit{cultural} context \citep{nowak_evolution_2005, alexander_biology_1987}. For instance, sharing rates in DG vary widely across societies, which in turn has been ascribed to variation in fairness norms \cite{henrich_search_2001, henrich_weirdest_2010, kawamura_altruism_2020}.

Similarly to variation in norms regarding the appropriate behavior in lab experiments, there is variability in norms regarding \textit{real-world prosocial behaviors}. For example, cultural variation exists in need-based communal norms versus conditional exchange norms \citep{miller_cultural_2017} and in prosocial norms related to environmentalism \citep{torgler_environmental_2009}. Moreover, cross-cultural research found significant variation in the \textit{performance} of prosocial behaviors across societies, such as variation in rates of volunteering \cite{ruiter_national_2006} and charitable giving \cite{bauer2013time}. However, the relationship between \textit{real-world} prosocial behavior and norms \cite[as has been established for lab experiments by][]{krupka2013identifying} remains unclear.

Thus there exists well-documented contextual variation in (prosocial) norms, and we argue that norms applying to the provision of incentives for behaving prosocially are no exception. That is, we deem it plausible that norms vary across incentive types and cultural contexts, and that this has reputational effects which in turn influence behavior. Akin to experimental investigations of the effects of norms on prosocial behavior \cite{krupka2013identifying}, we aimed to test the relationship between norms (regarding incentives) and \textit{real-world} prosocial behavior. Instead of simply assuming that people want to be perceived as prosocial or as intrinsically motivated as possible \cite[as formalized for example in][]{benabou_incentives_2006}, we propose to explicitly take social norms into account. 

\paragraph*{The current research.}

In the following, we detail a theoretical model of prosocial behavior that posits a crucial role of social norms in determining the effects of incentives on prosociality. We empirically test the model on a real-world prosocial behavior: blood donation. Ensuring a sufficient supply of blood is very important for all societies, as blood plays an essential part in many medical procedures (e.g., surgeries, cancer treatment) and for producing life-saving drugs. The case of blood donation is also theoretically interesting, as there have been many debates about whether donors should be offered incentives \citep{titmuss_gift_1971}, and as a result different countries have adopted various policies regarding incentives for donors \citep{chell_systematic_2018, healy_embedded_2000}. 

\section*{Results}

\paragraph*{A formal model of prosocial behavior.}

We implement a formal model of prosocial behavior based on the prominent model of prosocial behavior by Bénabou and Tirole \cite{benabou_incentives_2006} (henceforth BT). We follow BT in positing that prosocial behavior is governed by (a) the direct benefit and (b) the reputational benefit associated with performing the behavior. Specifically, an agent A decides whether to engage in the prosocial behavior $B$ by maximizing his benefits: 

\begin{equation} 
\label{eqn:max}
 Max		\{ \mbox{\textit{direct benefit + reputational benefit}} \}, B \in \{0,1\}
\end{equation}

Here, \textit{direct benefit} is itself composed of three subcomponents: intrinsic motivation $v_{a} \in [0,1]$, extrinsic motivation $v_{v} \in [0,1]$ and cost $c \in [0,1]$ (e.g., in terms of time or resources). The latter has a negative effect, whereas the former two have positive effects on motivation to perform $B$. Importantly, extrinsic motivation $v_{v}$ can play a role only if an incentive $R$ is provided (i.e., if $R = 1$; $R \in \{0,1\}$), otherwise $v_{v} = 0$. \textit{Direct benefit} is thus defined as follows:

\begin{equation} 
\label{eqn:direct_benefit}
 \mbox{\textit{direct benefit}} = B * (v_{a} + v_{v} * R) - B * c
\end{equation}

\textit{Reputational benefit} represents the benefit (or cost) of being perceived in a certain way by others, i.e., the reputational outcomes related to the behavior. BT theorized \textit{reputational benefit} as depending on $VIS$, the visibility of the action (i.e., to what extent the behavior is performed in private vs. public), as well as $pref_{v_{a}}$ and $pref_{v_{v}}$, the individual’s preference for appearing intrinsically and not extrinsically motivated respectively. Most importantly, reputational motivation depends on the \textit{expected value} of an individual’s intrinsic and extrinsic motivation given their behavior and the extrinsic reward associated with the behavior. Crucially, BT assumed that the expected value of extrinsic motivation $E(v_{v})$ is a \textit{cost} whereas the expected value of intrinsic motivation $E(v_{a})$ is a \textit{benefit}. 

Notably, expected intrinsic motivation $E(v_{a} | R, B)$ as formulated in the model has a positive effect irrespective of whether incentives are offered. Individuals who engage in the prosocial activity have a higher expected intrinsic motivation than individuals who do not behave prosocially. However, the same does not hold for the cost associated with higher expected extrinsic motivation. In the presence of incentives, individuals engaging in the prosocial activity are perceived to have higher extrinsic motivation than individuals not engaging in the prosocial activity. The model thus demonstrates that "motivational crowding-out" can arise endogenously, as the cost of being perceived as extrinsically motivated\footnote{In addition to a reputational cost from appearing extrinsically motivated, decreased perceived \textit{intrinsic} motivation as a result of receiving an incentive presents a further, albeit minor, reputational cost. This is because $E(v_{a} | R = 0, B = 1)$ is slightly larger than $E(v_{a} | R = 1, B = 1)$; the extent of this reputational cost depends on the costliness of performing the prosocial behavior (see SI Appendix Fig. S2).} offsets the benefits of extrinsic motivation itself. 

In the following, we deviate from BT's model by introducing two additional parameters capturing social norms. We replace the hard-coded cost (benefit) to being perceived as extrinsically (intrinsically) motivated with two social norm variables in order to account for contextual variation in how acceptable members of society view appearing intrinsically or extrinsically motivated by a given incentive offered for a particular prosocial act. We call this social norm variable $S$ and define it as the \textit{sum of reputational costs and benefits as dictated by the prevalent social norms associated with a given behavior}. This applies to both expected intrinsic motivation (i.e., $S_{v_{a}} \in [-1,1]$) and expected extrinsic motivation (i.e., $S_{v_{v}} \in [-1,1]$). From this follows our definition of reputational benefit:

\begin{equation} 
\begin{split}
 \mbox{\textit{reputational benefit}} = VIS[\mathbin{\color{red}\textbf{$S_{v_{a}}$}} * pref_{v_{a}} * E(v_{a} | R, B) +\\ \mathbin{\color{red}\textbf{$S_{v_{v}}$}} * pref_{v_{v}} * E(v_{v} | R, B)]
 \end{split}
\end{equation}

\begin{figure}[t]
\centering
\includegraphics[width=\linewidth]{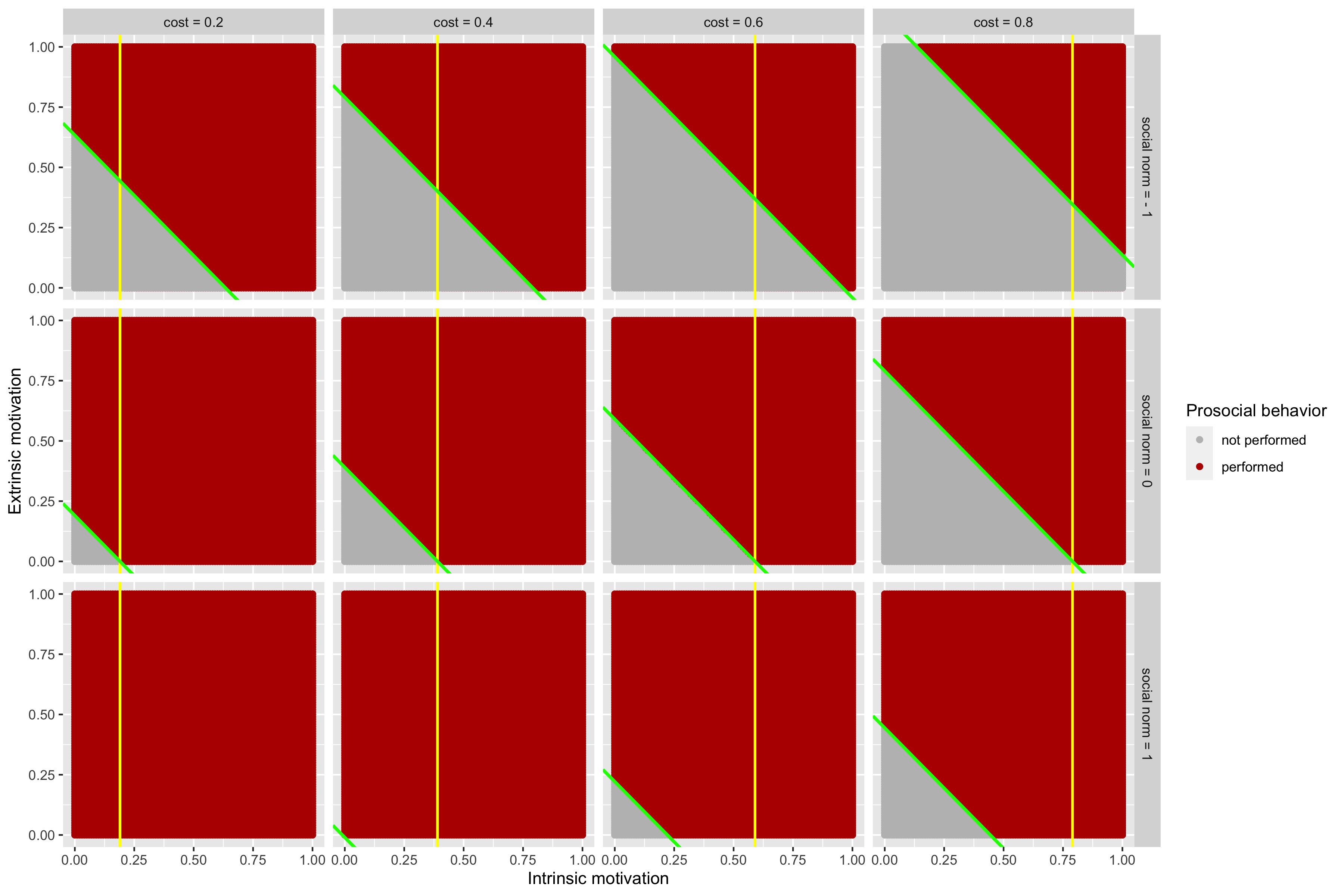}
\caption{Simulated behavior of individuals differing in intrinsic (x-axis) and extrinsic (y-axis) motivation for different values of social norms (rows) and cost (columns) when incentives are offered. In the presence of incentives, the value of the social norms variable $S_{v_{v}}$ determines the shift on the x-axis of the diagonal line dividing those behaving prosocially and those not (green line; see description in main text). Parameter settings: $VIS = pref_{v_{a}} = pref_{v_{v}} = 1$; $S_{v_{a}} = 0$; $v_{a} \sim unif\{0,1\}$ and $v_{v} \sim unif\{0,1\}$.}
\label{fig:model}
\end{figure}

To illustrate the effect of our model modification, let us consider the effect of varying levels of $S_{v_{v}}$ on prosocial behavior in the presence of incentives. When $S_{v_{v}}$ takes on the value $0$, as depicted in the second row of Fig. \ref{fig:model}, a prosocial act is performed if the additive effects of intrinsic and extrinsic motivation outweigh the costs. That is, individuals whose \textit{intrinsic} motivation is larger than the costs of displaying the behavior -- who are located to the right of the yellow line -- perform the prosocial act. And in addition, individuals with insufficient intrinsic, but some \textit{extrinsic} motivation -- who are located between the yellow and green diagonal lines -- are also motivated to display the behavior. However, when $S_{v_{v}}$ takes on a \textit{negative} value (Fig. \ref{fig:model}, first row), the diagonal threshold dividing those behaving prosocially and those not (green line) shifts toward the right on the x-axis. This is due to agents incurring a reputational cost arising from social norms that characterize it as “bad” that the given prosocial act is performed while the agents receive a specific incentive\footnote{Note that we take \textit{social norms regarding incentive X} as a proxy for our theoretical construct \textit{social norms for being extrinsically motivated by incentive X} ($S_{v_{v}}$), because the main reputational consequence of introducing incentives is that individuals appear more extrinsically motivated (see SI Appendix Fig. S2).}. On the other hand, when $S_{v_{v}}$ is positive (Fig. \ref{fig:model}, third row), the green threshold shifts to the left on the x-axis, reflecting high levels of prosocial behavior due to the combined positive effect of intrinsic, extrinsic and reputational motivation.

\paragraph*{Testing the model on a real-world prosocial behavior.}

\begin{figure}
\centering
\includegraphics[width=.9\linewidth]{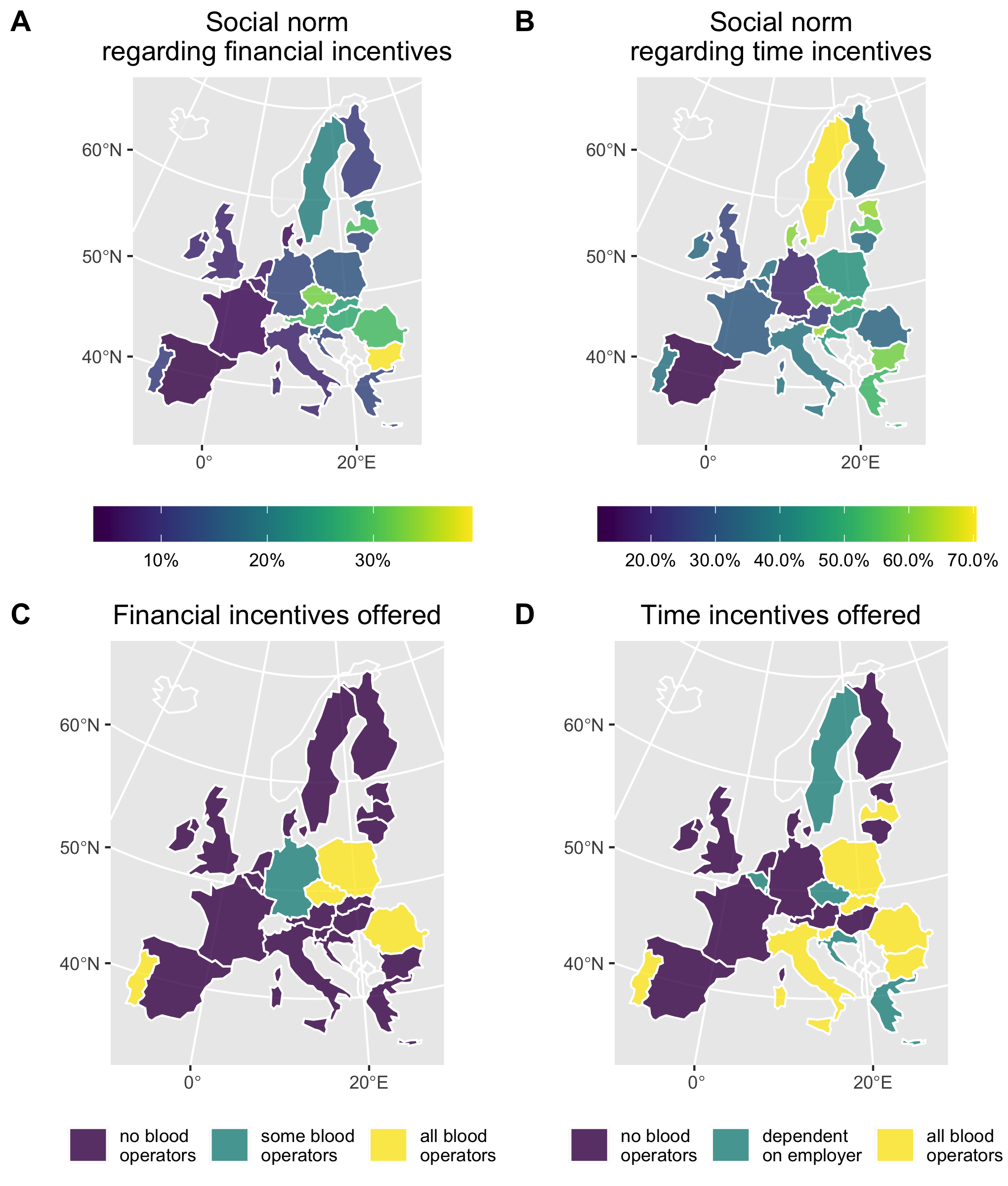}
\caption{Maps displaying blood donor incentives and norms regarding the acceptability of incentives across Europe. \textbf{(A)} Social norm regarding the acceptability of \textit{financial} incentives for blood donation. \textbf{(B)} Social norm regarding the acceptability of \textit{time} incentives for blood donation. \textbf{(C)} Financial incentives and \textbf{(D)} time incentives offered to blood donors (where the incentive may be provided by \textit{all}, only \textit{some} or \textit{none} of the blood operators in a given country; in the case of \textit{time} incentives, the incentive may be offered only to \textit{donors whose employer allows it}).}
\label{fig:maps}
\end{figure}

We test our formal model empirically on the real-world prosocial behavior of blood donation\footnote{Specifically, we are interested in examining how prosocial behavior is related to incentives and social norms, as well as to intrinsic and extrinsic motivation. Thus, our focus is on the model parameters $v_{a}$, $v_{v}$, $c$, $R$ and $S_{v_{v}}$. The effects of other model parameters on prosociality are documented elsewhere \cite[e.g., visibility][]{ariely_doing_2009}.}. To this end we employed large-scale survey data from the Eurobarometer \cite{european_commission_eurobarometer_2014}, which was obtained from representative samples of 28 European countries (n = 26 192). Overall, 38.4\% (n = 10195) of all respondents included had ever donated blood. This varied substantially across countries (see SI Appendix Fig. S3). Similarly, we found strong variation in social norms regarding incentives (see Fig. \ref{fig:maps}A and B; social norms regarding financial incentives: ${X}^2(27) = 1970.7, p < 0.001$; social norms regarding time incentives: ${X}^2(27) = 2436.6, p < 0.001$). However, social norms regarding \textit{time} incentives were more positive (mean = 0.43) than those regarding \textit{financial} incentives (mean = 0.15; Mann–Whitney $U(n_{1} = n_{2} = 28) = 736, p < 0.001$). Moreover, both financial and time incentives offered to blood donors varied across countries (see \ref{fig:maps}C and D). See SI Appendix for more descriptive information and figures. 

\begin{figure*}[b!]
\centering
\includegraphics[width=\textwidth]{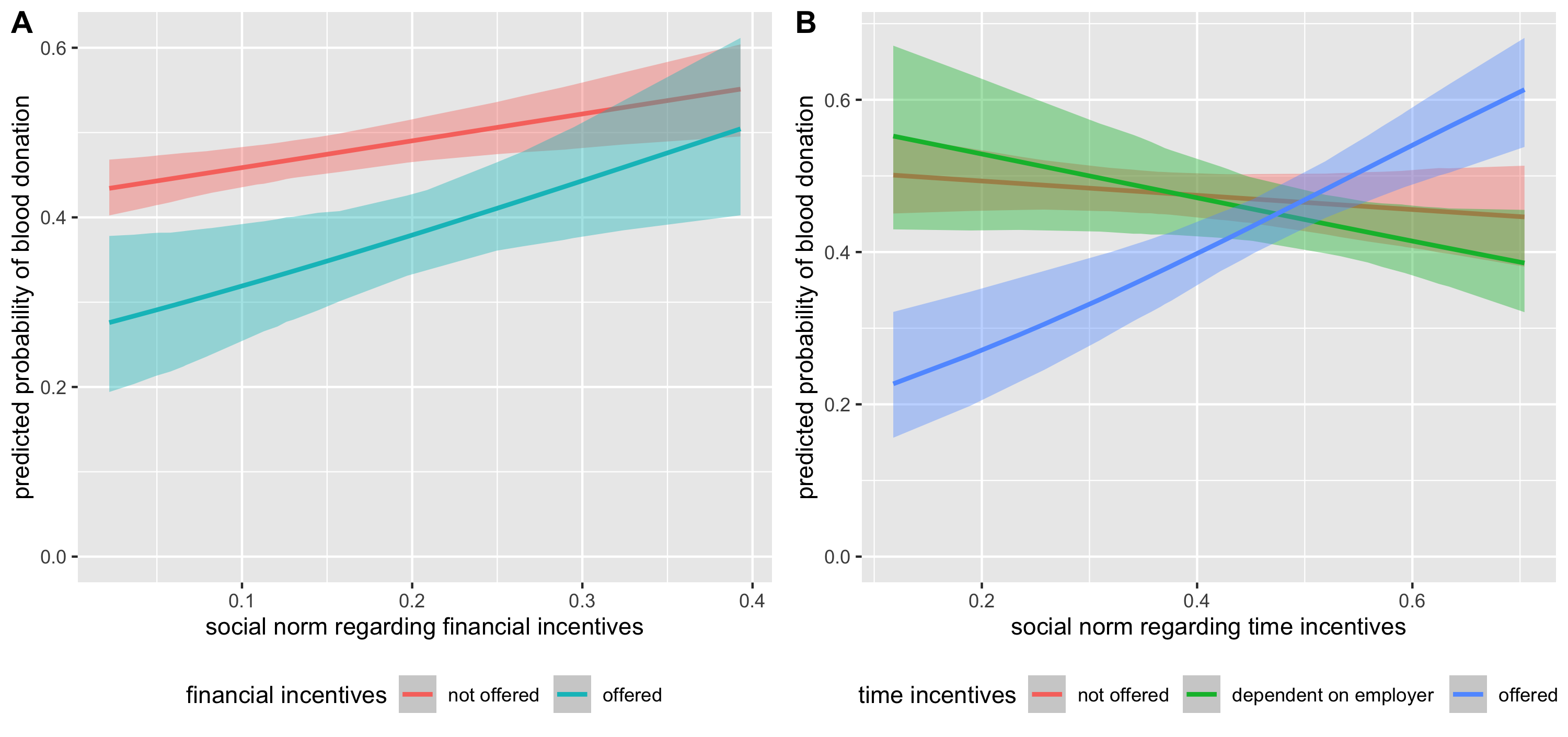}
\caption{Predictive margins of donating blood for a representative European woman (i.e., mean-aged, living with a partner in a small town, employed, having completed secondary-level education, mean number of children, rarely having difficulty paying bills, intrinsically but not extrinsically motivated) as a function of \textbf{(A)} social norms regarding \textit{financial} incentives, split by financial incentives (where incentives are offered to all blood donors in a given country or not at all) and \textbf{(B)} social norms regarding \textit{time} incentives, split by time incentives (where incentives are offered to all blood donors in a given country, only to donors whose employer allows it, or not at all). Prediction bands indicate 80\% prediction intervals.}
\label{fig:main_results}
\end{figure*}

We found a positive main effect of social norms regarding \textit{financial} incentives on blood donation propensity ($b = 0.125, p < 0.05$), but a negative main effect of financial incentives themselves ($b = -0.512, p < 0.01$). As illustrated in Fig. \ref{fig:main_results}A, these results indicate that respondents from countries where incentives are offered are, overall, less likely to have donated blood. Yet, the likelihood of having donated was higher in countries with more positive norms regarding financial incentives. We did not find a significant interaction effect between financial incentives and social norms. However, we did find a positive interaction between \textit{time} incentives and social norms ($b = 0.484, p < 0.01$). That is, people living in a country providing time incentives are more likely to donate blood when social norms regarding time incentives are more positive (see \ref{fig:main_results}B). 

In addition, the full model revealed a positive effect of self-reported intrinsic motivation ($b = 1.346, p < 0.0001$), and a negative effect of cost of donating (in terms of living in rural areas) on giving blood ($b = -0.091, p < 0.05$). Individuals who reported being motivated by the desire to help those in need, alleviate shortages or support medical research and those living in urban areas are more likely to have donated blood\footnote{As a robustness check, we implemented the same model on a subset of respondents, excluding those who indicated they are not willing to donate in the future (these respondents were not asked about motivational factors; see SI Appendix). For this sample we also found a positive, albeit smaller, effect of intrinsic motivation on blood donation ($b = 0.091, p < 0.05$).} (see SI Appendix Tab. S3 and S4 for full model results). The effect of self-reported extrinsic motivation (in terms of getting something in return as a motive to donate) was in the expected direction but non-significant ($b = 0.110, p = 0.078$).

\section*{Discussion}

In this paper we extended a model of prosocial behavior that posits an additive effect of intrinsic motivation, extrinsic motivation, reputational motivation, and costs. Our key theoretical contribution is the introduction of social norms. We argue that reputational costs or benefits are a product of social norms, i.e., how acceptable certain behaviors are viewed by members of society. In our view this theoretical modification allows us to account for the inconsistent effects of incentives on prosocial behavior across contexts. We tested the model on a real-world prosocial behavior, namely blood donation, and found a positive association between country-level social norms for financial incentives and blood donation: the more positive the social norm, the higher the probability of having donated blood. While financial incentives were negatively associated with blood donation, time incentives were on their own unrelated to the likelihood of donating. Instead, the effect of time incentives depended on social norms; i.e., in countries where time incentives are offered, individuals’ likelihood of donating was higher when social norms regarding these incentives were positive than when social norms were more negative.

Akin to previous studies \cite[see ][]{gneezy_when_2011, chell_systematic_2018}, we found that incentives have inconsistent effects on prosocial behavior. Whereas time incentives alone were unrelated to an individual's likelihood of having donated blood, financial incentives were negatively associated with the propensity to donate blood. However, the differences in the relationship between financial and non-financial incentives and prosociality can in fact be explained by norms -- i.e., the social norm regarding the incentive predicts whether the probability of donating increases. On the one hand, time incentives were viewed rather negatively in some of the countries offering them, and respondents from these countries were less likely to donate compared to respondents in countries without time incentives. On the other hand, respondents in countries where social norms regarding time incentives were more positive were \textit{more} likely to donate than comparable respondents from countries not offering time incentives. Similarly for financial incentives, the likelihood of donating increased if social norms were more positive, even though financial incentives themselves had a negative effect on the likelihood of donating blood. As such, our findings highlight the relevance of BT's theory in a large population and for a real-world prosocial behavior. But more importantly, our conclusions suggest that our theoretical modification derived from interdisciplinary cross-pollination -- the integration of social norms -- provides a fruitful addition to our understanding of the relationship between incentives and prosocial behavior.

While we found that more positive norms regarding \textit{time} incentives are associated with a higher predicted probability of donating \textit{if} time incentives are offered, our observation that social norms regarding \textit{financial} incentives are positively associated with donation behavior \textit{irrespective of whether incentives are offered} is unexpected. A possible explanation for this may be that more positive norms regarding financial incentives capture additional societal attitudes that impact donation behavior, e.g., valuation of the donor status generally. 

The most direct implication that follows from our findings is that incentives \textit{can} be associated with increased levels of prosocial behavior -- \textit{if} social norms regarding the incentive are sufficiently positive. Although clever experimental interventions have been designed to avert the reputational \textit{costs} arising from offering incentives \cite[e.g.,][]{kirgios2020forgoing}, our study suggests that there need not be such \textit{costs} for all combinations of incentives and prosocial activities, but that reputational consequences depend on the relevant norms. Thus, experimenting with different types of incentives (which may be perceived as more or less appropriate in a given society) seems promising as a way to find incentives that encourage societally desirable behavior with less or no reputational cost. In addition, it may be possible to increase the effectiveness of incentives by changing how incentives are framed.  For instance, blood operators might be able to promote views emphasizing how fair it is that donors receive time off work due to their critical role in the medical system.

While our findings entail important theoretical and policy implications, open questions remain for future research to explore. Even though we found that a relatively simple measure of social norms explains interesting empirical patterns of real-world prosocial behavior, a more fine-grained measure of social norms \cite[e.g., according to][]{rauhut_sociological_2010, krupka2013identifying} would offer further important insights. For instance, a measure targeting more specifically the acceptability of \textit{being extrinsically motivated} by incentive X to perform behavior Y could reveal precisely which norms govern the reputational consequences of receiving an incentive for behaving prosocially. In addition, the operationalization of intrinsic and extrinsic motivation could be improved, as the Eurobarometer employed a selection criterion for these measures (see SI Appendix). Although intrinsic and extrinsic motivation were not our main concern in this paper, further examination of the interplay of various motivational factors and social norms seems promising. Lastly, we cannot draw any causal conclusions from our observational methods. Our finding of lower donation rates in countries offering financial incentives does not necessarily mean that incentives do not encourage donation, and the association between country-level social norms and donation rates may be due to other factors. Experimental investigations could further our understanding of the effects of norms on prosocial behavior in the presence of incentives, including to what extent norms can be manipulated in the lab and field. 

Nevertheless, our unique approach allowed us to investigate the effect of social norms “in the wild”; That is, we were able to examine empirically elicited norms across countries which are typically very persistent and have a long cultural evolutionary history. Our observation that these norms offer an explanation both for cross-country variation and differences across types of incentives highlights not only the theoretical merits of incorporating social norms, but also offers concrete implications for policy-makers regarding how to encourage and influence prosocial behavior more generally.

\section*{Materials and Methods}

All data and code for running the formal model and analyses, as well as our preregistration\footnote{Note that only analyses pertaining to donation of \textit{blood} (not other substances of human origin) are presented here. We deviate from the preregistration in two respects. Firstly, instead of creating a “conditional” social norm and extrinsic motivation variable (i.e., setting these variables to 0 if no incentive is offered), we employ interaction terms between incentives and social norms / extrinsic motivation. Secondly, we did not run a robustness check excluding those countries not offering incentives, because this would have resulted in too small a sample size at the country-level.}, are available at https://osf.io/bf2kt/.

\paragraph*{Participants and procedure.}
We employed comparative data from the 2014 wave of the Eurobarometer \cite{european_commission_eurobarometer_2014}, which is a repeated cross-sectional survey conducted on representative samples from European Union member states. In 2014, the survey was performed in 28 countries. A multi-stage random (probability) sampling design was employed and approximately 1,000 face-to-face interviews were completed in each of the countries. Respondents were residents in the respective country, had sufficient command of the national language(s), and were 15 years or older. Study protocols were approved by the European Commission and all participants provided informed consent. A detailed description of all variables, including the wording of survey items and measurement scales, is presented in SI Appendix, Tab. S5.

\paragraph*{Blood donation.}

The dependent variable in our analyses is blood donation during one's lifetime (yes/no). Participants in the Eurobarometer were asked whether they have ever donated substances of human origin, including blood. All categorical variables, including the dependent variable, are dummy-coded.

\paragraph*{Incentives.} Based on responses to the expert survey (see below), we constructed two country-level incentive variables: (a) financial incentives (i.e., cash or other high-value incentives, for example tax deduction) and (b) time incentives (i.e., time off work). There were three categories of financial incentives: (1) offered by \textit{no} blood operators, (2) \textit{some} blood operators, and (3) all blood operators. Time incentives also had three levels: (1) offered by \textit{no} blood operators, (2) offered \textit{dependent on the employer} (i.e., in some countries employers could grant donors permissions to donate during working hours, but this did not hold for all donors), and (3) offered by all blood operators.  

\paragraph*{Social norms regarding incentives.} As social norms by definition apply to the (societal) group level, we operationalized \textit{country-level} acceptability ratings of incentives rather than using respondents’ individual attitudes. Specifically, we aggregated the responses of individual respondents regarding the acceptability of receiving incentives for blood donation in each of the 28 European countries. This resulted in a country-level variable which ranges from 0 (i.e., 0\% of respondents from a given country find financial/time incentives acceptable) to 1 (i.e., 100\% of respondents from a given country find financial/time incentives acceptable). We computed social norms separately for financial and time incentives. 

\paragraph*{Intrinsic and extrinsic motivation.} Intrinsic motivation is operationalized as the mentioning (yes/no) of an intrinsic motivational factor as a reason for donating blood. Intrinsic motivation is coded “yes” if one of three intrinsic motivational factors are mentioned: (a) to help other people in need, (b) to alleviate shortages of blood, or (c) to support medical research. Extrinsic motivation both for financial and time incentives is operationalized as the mentioning (yes/no) of an extrinsic factor as a reason for donating or being willing to donate blood (namely “to receive something in return for you or your relatives”). We additionally included two other measures of extrinsic motivation for \textit{financial} incentives: (1) subjective difficulty in paying bills (three levels: almost never/never, from time to time, most of the time; we assumed higher extrinsic motivation would follow from increased difficulty in paying bills) and (2) lack of current employment status (not employed/employed; we assumed higher extrinsic motivation for \textit{financial} incentives would follow from being \textit{un}employed). For time incentives, we included the presence of employment as an extrinsic motivator (employed/not employed;  we assumed extrinsic motivation for \textit{time} incentives would be higher for those employed).

\paragraph*{Additional covariates.} We controlled for socio-demographic factors, which previous studies had shown to be related to propensity to donate (see \cite{piersma_individual_2017} for a recent review on donor demographics). We included (a) age (as our dependent variable is dependent on years lived), (b) gender (male/female), (c) education (in years, categorized in four levels), and (d) partner status (cohabitating: yes/no). In addition, we controlled for two factors which represent costs of donating: (e) time constraints due to childcare responsibilities (operationalized as number of children in household), and (f) travel costs due to living in rural areas (i.e., rural areas have fewer blood collection locations; three levels: large town; small/middle-sized town; rural area).

\paragraph*{Expert survey.}
We conducted a survey of incentive policies among specialists at national blood collection agencies across Europe. Experts reported on the type and value of incentives provided in their country by means of a telephone interview or in writing. We explicitly asked experts not to mention low-value incentives (as it is common for donors to receive small gifts such as T-shirts or mugs); we are interested only in high-value (non-symbolic) incentives. Incentive data for the Czech Republic, France, and the Netherlands were taken from a recently published study on donor incentives across the globe \citep{zeller_vox_2020}.

\paragraph*{Analytical strategy.}
Respondents younger than 18 years were excluded since most EU countries do not permit people under 18 to donate blood\footnote{In addition, respondents from Germany were excluded from analyses pertaining to financial incentives since Germany was the only country where blood operators across the country differed in the financial incentives provided. The results are quantitatively equivalent when Germany is included and regrouped either as offering, or not offering, financial incentives.}. We employed logistic multilevel mixed models for all main analyses, where individuals were nested in countries. The models included individual-level variables (e.g., intrinsic motivation) and country-level variables (e.g., incentives) as fixed effects, as well as allowing the intercept to vary by country. The estimation technique is maximum likelihood. Missing values were handled by list-wise deletion. We ran separate models for financial and time incentives. The predictors of interest are intrinsic motivation, cost, extrinsic motivation in interaction with financial/time incentives, and social norms regarding financial/time incentives in interaction with financial/time incentives. Prediction intervals are based on fixed coefficients and observation-level errors. All analyses were run with R version 4.0.2 using the packages lme4 (modeling), merTools (predictions), tidyverse (data transformation; plotting) and rnaturalearth (maps) \cite{bates2007lme4, knowles2016package, wickham2019welcome, south2017rnaturalearth}. 

\bibliographystyle{plainnat}
\bibliography{main.bib}

\newpage

\pagenumbering{roman}

\newgeometry{total={6.7in, 9in}}

\centerline{\textbf{\LARGE{Appendix}}}

\renewcommand{\figurename}{Figure S}
\setcounter{figure}{0}

\section*{Supplementary results}

\subsection*{Theoretical model: Direct benefit}

As described in the main text, we follow Bénabou and Tirole \cite[][henceforth BT]{benabou_incentives_2006} and assume that the \textit{direct benefit} gained from performing a prosocial behavior $B$ is dependent on an individual's intrinsic motivation $v_{a}$, extrinsic motivation $v_{v}$, as well as the costs $c$ and incentive $R$ associated with performing the behavior. We defined direct benefit as follows: 

\begin{equation} 
 \mbox{\textit{direct benefit}} = B * (v_{a} + v_{v} * R) - B * c
\end{equation}

In the absence of extrinsic rewards (i.e., $R = 0$), only intrinsic motivation $v_{a}$ and personal cost $c$ determine whether the prosocial action is performed. Personal cost here refers to the individual-level cost associated with the activity, for instance in terms of time, money, and resources. Although costliness of the behavior varies across individuals, it is also conceivable that some activities are highly costly for many people (such as donating a kidney) and are thus performed only by very few who have exceptionally high intrinsic motivation and relatively low cost. On the other hand, some behaviors may be associated with relatively low costs for most people (such as helping someone who tripped and fell) and are therefore performed by most people who have sufficiently high intrinsic motivation. Assuming that the prosocial behavior is performed if \textit{direct benefit} is greater for $B = 1$ than for $B = 0$, the effect of cost\footnote{For illustrative purposes, cost is here assumed to be the same for all individuals. This is however not a constraint in our model, where cost is able to vary across people.} is illustrated in Fig. S \ref{fig:model_IM_EM}: The first row depicts the effect of varying the cost of the prosocial activity in the absence of incentives. In the upper leftmost facet, most people engage in the behavior (red), because relatively low levels of intrinsic motivation suffice to engage in the relatively uncostly behavior. On the upper rightmost facet is the other extreme, where only the most highly intrinsically motivated individuals perform the prosocial action given the high associated costs.

When incentives are offered (i.e., $R = 1$), the decision to perform the prosocial behavior is governed both by extrinsic and intrinsic motivation as well as costs. This results in a higher total number of individuals that join in performing the behavior, because some individuals are motivated by the extrinsic reward. The bottom row in Fig. S \ref{fig:model_IM_EM} illustrates this effect: the dividing line between those who engage in the prosocial behavior and those who do not virtually tips over toward the left side, as individuals with insufficient intrinsic motivation but some extrinsic motivation now choose to perform the behavior as well. 

\subsection*{Theoretical model: Reputational benefit}

BT \cite{benabou_incentives_2006} further postulate that \textit{reputational motivation} influences the decision to engage in a prosocial activity or not. As described in the main text, reputational motivation is dependent on an agent's expected extrinsic motivation $E(v_{v})$ and expected intrinsic motivation $E(v_{a})$ given the rewards offered (where the former represents a cost and the latter represents a benefit for the agent's reputation), as well as $pref_{v_{a}}$ and $pref_{v_{v}}$ (i.e., the strength of preference to appear intrinsically motivated and not extrinsically motivated) and $VIS$ (the visibility of the action). Reputational motivation according to BT \cite{benabou_incentives_2006} is then defined as follows:

\begin{equation} 
 \mbox{\textit{reputational benefit}} =  VIS[pref_{v_{a}} * E(v_{a} | R, B) - pref_{v_{v}} * E(v_{v} | R, B)]
\end{equation}

As explained in the main text, the first component of \textit{reputational benefit} (namely $VIS * pref_{v_{a}} * E(v_{a} | R, B)$; of which the driving factor is the expected value of intrinsic motivation $E(v_{a} | R, B))$ always has a positive effect: Individuals who engage in the prosocial activity have a higher expected intrinsic motivation than individuals who do not behave prosocially, irrespective of whether an extrinsic reward is offered. That is,

\begin{equation} 
\label{eqn:SI1}
 E(v_{a} | R = 0, B = 1) \mathbin{\color{red}>} E(v_{a} | R = 0, B = 0)
\end{equation}
\centerline{and}
\begin{equation} 
\label{eqn:SI2}
 E(v_{a} | R = 1, B = 1) \mathbin{\color{red}>} E(v_{a} | R = 1, B = 0)
\end{equation}

Yet, the same does not hold for the cost associated with higher expected extrinsic motivation: In the presence of incentives, the expected value for extrinsic motivation is in fact higher for individuals engaging in the prosocial activity than not, and this is not the case when there are no incentives provided:

\begin{equation} 
\label{eqn:SI3}
 E(v_{v} | R = 0, B = 1) \mathbin{\color{red}=} E(v_{v} | R = 0, B = 0)
\end{equation}
\centerline{and}
\begin{equation} 
\label{eqn:SI4}
 E(v_{v} | R = 1, B = 1) \mathbin{\color{red}>} E(v_{v} | R = 1, B = 0) 
\end{equation}

Increased levels of expected extrinsic motivation when incentives are present, combined with ascribing a cost to being perceived as extrinsically motivated, thus introduces a key negative reputational consequence resulting from offering incentives for a prosocial behavior. In addition to this reputational cost from increased \textit{expected extrinsic motivation}, there is a minor reputational cost resulting from decreased \textit{expected intrinsic motivation} associated with performing a prosocial behavior for which an incentive is provided, compared to performing a prosocial behavior for which no incentive is offered. That is, the expected intrinsic motivation of an agent performing a prosocial act is always higher than that of another agent not performing the behavior, but the \textit{extent} of this reputational benefit depends on whether an incentive is offered: Expected intrinsic motivation when performing a prosocial act for which an incentive is offered (i.e., $E(IM | R = 1, B = 1)$) is lower than the expected intrinsic motivation when performing a prosocial act for which \textit{no} incentive is offered (i.e., $E(IM | R = 0, B = 1)$). The difference in expected intrinsic motivation for these two scenarios is dependent on costliness of performing the behavior (see Fig. S \ref{fig:model_rep}, blue line), such that the relative reputational cost from expected intrinsic motivation is higher the lower the costliness of the prosocial behavior. Similarly, the size of the reputational cost due to expected extrinsic motivation is also dependent on costliness of the prosocial behavior (see Fig. S \ref{fig:model_rep}, red line). While the reputational costs induced by incentives in terms of expected \textit{intrinsic} motivation increase as costliness of performing the prosocial act increases, the costs induced by incentives in terms of expected \textit{extrinsic} motivation decrease with increasing costliness of performing the prosocial act.

In sum, the assumptions by BT \cite{benabou_incentives_2006} that (a) reputational \textit{costs} are associated with being perceived as extrinsically motivated and (b) reputational \textit{benefits} are associated with being perceived as intrinsically motivated, result in reputational \textit{costs} for agents who perform a prosocial behavior in the presence of incentives. These reputational costs induced by incentives are largely driven by costs in terms of being perceived as more extrinsically motivated, but there is also a minor cost due to being perceived as less intrinsically motivated.

\subsection*{Empirical results: Descriptives}

Donation rates ranged from 22.8\% of respondents from Portugal having ever donated blood to 52.9\% of respondents from France having donated blood (see Fig. S \ref{fig:map_BD}). Table S \ref{table:descriptives} displays descriptives of demographic variables and predictors of interest. The correlation matrix of predictors of interest is visualized in S \ref{fig:corr_plot}. Figure S \ref{fig:map_IM_EM} further displays the geographic distribution of country-level mean levels of intrinsic and extrinsic motivation, which also showed substantial variation (intrinsic motivation: ${X}^2(27) = 1501.5, p < 0.001$; extrinsic motivation: ${X}^2(27)= 408.58, p < 0.001$). Notably, levels of intrinsic motivation to donate (see S \ref{fig:map_IM_EM}A) were high across all countries (country-level means ranging from 41.3\% in Bulgaria to 88.0\% in Sweden), albeit higher in Western and Northern European countries compared to Eastern European countries. In contrast, mentioning the extrinsic factor “getting something in return” as motivation to donate was relatively rare (see S \ref{fig:map_IM_EM}B; country-level means range from 1.3\% in Italy to 12.7\% in Finland).

\paragraph{Incentives}

The expert survey revealed that only few European countries are offering high-value \textit{financial} incentives to blood donors: In Bulgaria, the Czech Republic, Poland, and Portugal all blood operators offer high-value incentives to donors. In Germany, some blood operators offer incentives (e.g., hospital-based blood operators), whereas others (e.g., the Red Cross) do not. \textit{Non-financial time} incentives, however, are more commonly used to incentivize blood donation across Europe: 15 out of 28 countries offer some kind of time incentive. Countries that offer time off work to blood donors (independent of employer policies) include Eastern European countries such as Bulgaria and Romania, but also Southern European countries such as Italy and Portugal. In other European countries, a subset of employers allow donors to take time off work for making the donation (for instance in Belgium, Croatia, Greece and Sweden). Table S \ref{tab:incentives} provides detailed descriptions of financial and time incentives offered to blood donors across Europe.

\paragraph{Country-level rates of blood donation as a function of incentives and social norms}

Observational country-level mean blood donation rates as a function of the two country-level variables (a) incentives and (b) social norms are displayed in Fig. S \ref{fig:combined_descr}. Note that these are only raw observations, where no confounding factors are controlled for.

\subsection*{Empirical results: Statistical models}

Full model results (including results from robustness checks) are displayed in Tab. S \ref{tab:model_financial} (financial incentives) and Tab. S \ref{tab:model_time} (time incentives).

\paragraph{Country-level variation}

Models included random intercepts for country, which a log-likelihood test comparing the intercept-only model with and without random intercepts confirmed to be justified (${\chi}^2(1)=501.79, p < 0.001$).

\paragraph{Demographics}

Adding demographics to the model improved the model fit compared to the baseline (intercept-only) model further ($AIC_{demo} = 33097.4 < AIC_{intercept-only} = 34848.6$). As expected given that our dependent variable is blood donation \textit{during the lifetime}, older respondents were more likely to have donated ($b = 0.166, p < 0.001$). The other socio-demographic control variables were also significantly related to blood donation in the following manner: male ($b = -0.471, p < 0.001$) and more educated respondents (full-time education until 16-19 years: $b = 0.395, p < 0.01$; full-time education until 20 years or more: $b = 0.697, p < 0.001$), and those living with a partner ($b = 0.129, p < 0.001$) were more likely to have donated blood during their lifetime. Predictive margins of donating blood as a function of norms and incentives, split by gender, are displayed in Fig. S \ref{fig:pred_gender}.

\paragraph{Extrinsic motivation}

The effect of extrinsic motivation (in terms of getting something in return as a motivation to donate) was in the expected direction but non-significant ($b = 0.110, p = 0.078$ in models for financial incentives; $b = 0.104, p = 0.189$ in models for time incentives). Similarly, the interaction term between financial incentives and extrinsic motivation was in the predicted direction but did not reach significance (for financial incentive = 1: $b = 0.283, p = 0.09$; for time incentive = 1: $b = 0.108, p = 0.438$). In addition, the two other indicators with which we operationalized extrinsic motivation for \textit{financial} incentives (i.e., lack of employment and difficulty paying bills) had no positive effect on blood donation, contrary to our expectations. Instead, lack of employment was associated with \textit{lower} levels of blood donation and difficulty in paying bills was unrelated to donation behavior. In the models for \textit{time} incentives we included \textit{presence} of employment as an indicator of extrinsic motivation for time incentives, and indeed this had a positive effect on blood donation behavior ($b = 0.245, p < 0.001$).

\paragraph{Robustness checks}

We performed robustness checks for intrinsic and extrinsic motivation, because the Eurobarometer employed a selection criterion for questions regarding these two motivational variables: only respondents who had either donated in the past or were willing to donate in the future subsequently received the question concerning motivational factors. Since this selection criterion introduces significant bias which overestimates the effects of intrinsic and extrinsic motivation, we additionally ran all analyses only on those respondents who have donated blood or who indicated they are willing to donate in the future (i.e., excluding those indicating they are not willing to donate in the future). Results from these analyses are displayed in the rightmost columns in Table S \ref{tab:model_financial} (financial incentives) and S \ref{tab:model_time} (time incentives). 

Despite a sharp reduction in sample size and forgoing representativeness of the sample by excluding only a specific type of respondent, all effects found in the full model remained the same (except the demographic variable \textit{cohabitation}, which turned non-significant). However, as expected due to the selection criterion, the effect of intrinsic motivation became considerably smaller ($b = 0.091, p < 0.05$).

\subsection*{Exploratory analyses: Time incentives for employed and unemployed individuals}

We split the sample according to employment status and ran separate models for time incentives on these two subsets of respondents. While employed individuals should be motivated to receive a time incentive, those who are not employed cannot profit from time off work. However, the same results emerged for both of these models as for the main model for time incentives. This finding is likely due to temporal discrepancy of the measures (i.e., discrepancy between blood donation \textit{during the lifetime} and \textit{current} employment status; see limitations below).

\subsection*{Exploratory analyses: Region-level models}

Lastly, exploratory analyses which investigated the association between within-country regional social norms and blood donation behavior revealed qualitatively similar patterns of results to our main confirmatory analyses (see supp. Fig. S \ref{fig:region_scatter}A for financial incentives and S \ref{fig:region_scatter}B for time incentives). However, uncertainty around the predictions increases, indicating that differences across social norms at the within-country region level (where regions may reflect cultural or mere geographical clustering) may be more subtle and thus cannot explain behavior as well as the country-level social norm measures.

\section*{Supplementary discussion}

Obtaining (representative) samples from a sufficiently large number of countries is a major challenge for comparative studies such as ours. In the case of blood donation, the 2014 wave of the Eurobarometer is to our knowledge the only dataset which contains data on blood donation behavior that is comparable across multiple countries (as well as including attitudes towards different types of rewards and motivational factors). While this dataset thus has the advantage of offering rich population data for dozens of countries, our inability to precisely design the survey items has led to potential shortcomings regarding the operationalization of our theoretical constructs. 

One issue concerns the Eurobarometer employing a selection criterion for questions regarding intrinsic and extrinsic motivation: only respondents who had either donated in the past or were willing to donate in the future subsequently received the question concerning motivational factors. As our chief interest concerned the relationship between incentives, social norms and blood donation (for which data was available for all respondents), this issue does not compromise our main results, which were computed based on the full sample (including donors and non-donors). However, the selection criteria does result in bias when examining the effects of intrinsic and extrinsic motivation. For this reason we performed robustness checks by running the main analyses on the subset of participants who received the question related to motivational factors; However, these robustness checks suffer from their own shortcomings, for instance decreased power due to starkly reduced sample sizes. 

Moreover, it is problematic that the question regarding motivational factors included three intrinsic motivational factors as response options and only one extrinsic one, and that there may have been an order effect as the extrinsic motivational choice was always presented last. A measure of these motivational factors that targets more specifically individuals’ pleasure in helping others and their utility of (non-)monetary rewards may be better able to quantify the effects of intrinsic and extrinsic motivation. Additionally, the Eurobarometer permitted only a very simplistic measure of social norms, namely the aggregation of acceptability ratings at the societal level. Particularly when there is lack of consensus in a society regarding the acceptability of an incentive, uncertainty about negative consequences to his reputation may lead the respondent to self-report disapproval of financial incentives, especially because this dishonesty is not costly to him \cite[\textcolor{blue}{54},][]{rauhut_sociological_2010}.

Additionally, some of our predictors may be less reliable due to suffering from a temporal discrepancy. That is, whereas our dependent variable is blood donation \textit{during the lifetime}, many moderators are \textit{current} indicators at the time of survey completion, for instance occupation and financial situation. Thus the current situation of the respondent may not correspond to the situation during which he donated. As an example of why this may be problematic, take Germany: One major group of donors that makes use of monetary incentives in Germany are university students (\textcolor{blue}{55}), whose financial situation is relatively weak but who have good prospects of a high-paying position in the future. Lastly, we were not only constrained by the availability of \textit{indicators} in the survey, but also by the number of \textit{countries}: The inclusion of only 28 European countries limited the number of country-level indicators we could test in our multi-level models.

\section*{Supplementary methods}

\subsection*{Variable operationalization}

S \ref{tab:vars} provides an overview of all variables included in the statistical models, including a detailed description, wording of survey items, and the response range.

\subsection*{Exclusion of respondents}

For whole blood donation, 2.8\% of the participants did not answer the dependent variable question on donation behavior and were excluded from the sample. Moreover, we excluded respondents younger than 18 years old, as these individuals are not eligible to donate in most European countries; this led to an additional 2.0\% of respondents being excluded, resulting in a sample size of $n_{1} = 26,532$.

In addition, the model for financial incentives excludes respondents from Germany, as Germany was the only country where \textit{some} blood operators offer financial incentives. In all other countries financial incentives were offered either by all blood operators or by none. All results are quantitatively equivalent when Germany is included and regrouped either in the group of countries offering financial incentives or in the group not offering financial incentives. Note that for analyses pertaining to \textit{time} incentives Germany is included (as multiple countries fall into each of the time incentive categories), which results in differing sample sizes for these two analyses (i.e., for \textit{financial} incentives $n_{respondents} = 25068$, $n_{region} = 63$ and $n_{country} = 27$, whereas for \textit{time} incentives $n_{respondents} = 26532$, $n_{region} = 68$ and $n_{country} = 28$).

\subsection*{Analytic strategy for exploratory region-level models}

We followed the same analytic strategy for exploratory region-level analyses as we did for country-level analyses (i.e., including all individual-level variables such as intrinsic motivation, as well as \textit{incentive} as a country-level variable). The key difference in the region-level analyses was that social norms (both for financial and time incentives) were aggregated at the NUTS level 1 region-level instead of at the country-level. NUTS (Nomenclature of Territorial Units for Statistics) is a commonly used geocode standard for referencing subdivisions of countries. Hereby NUTS \textit{level 1} subdivisions are the next largest regional divisions after \textit{country}. In order to ensure that samples drawn from these regions are sufficiently representative, we excluded NUTS level 1 regions from which less than 100 respondents were drawn. This resulted in the exclusion of eight regions from five different countries; analyses were run on $n = 26005$ participants from 86 regions within Europe. We included \textit{regions} as additional random intercepts in all models. That is, the models assumed that participants are nested in regions, which again are nested in countries.


\afterpage{ \begin{figure}[p!]
    \centering
    \includegraphics[width=0.95\textwidth]{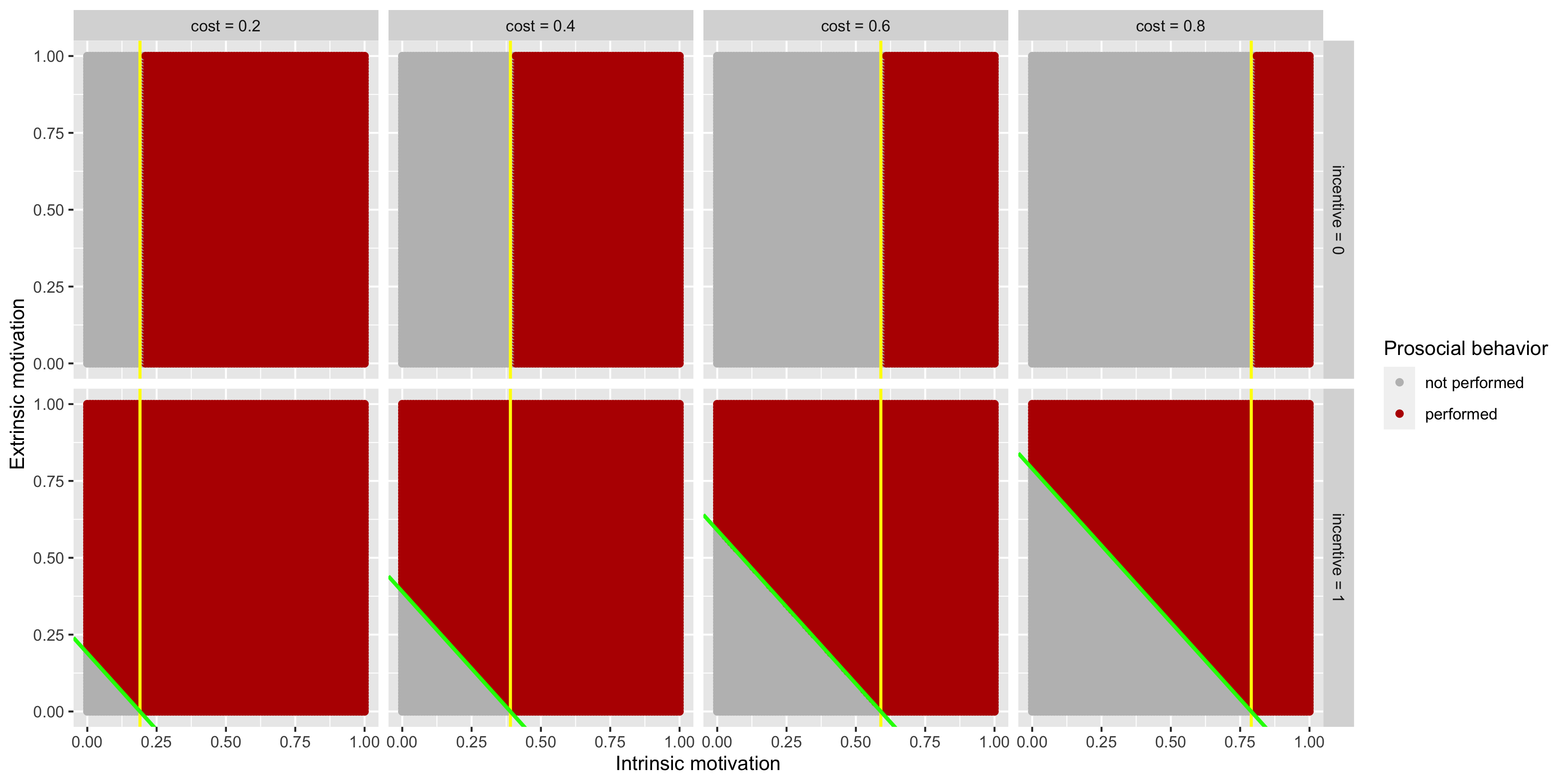}
    \caption{Visualization of the effects of varying cost of the behavior (as depicted in the four columns) and the presence of incentives (as depicted in the two rows) for individuals varying in intrinsic and extrinsic motivation (where $v_{a} \sim unif\{0,1\}$ and $v_{v} \sim unif\{0,1\}$). The first row illustrates that in the absence of extrinsic rewards, intrinsic motivation and cost govern whether a behavior is performed (yellow line). For example, if cost is 0.4 and incentive = 0, then those individuals whose intrinsic motivation is greater than 0.4 perform the prosocial behavior, irrespective of their extrinsic motivation. In the second row, where incentives are provided, the line dividing those taking the action versus not (green line) shifts counterclockwise: More people engage in the prosocial behavior, because sufficiently extrinsically motivated individuals also engage in the behavior. For example, if cost is 0.6 and incentive = 1, those individuals whose intrinsic motivation is greater than 0.6 perform the prosocial behavior, as in the first row, and \textit{additionally} those individuals whose sum of extrinsic and intrinsic motivation is greater than 0.6 perform the prosocial behavior.}
    \label{fig:model_IM_EM}
\end{figure} \clearpage }

\afterpage{ \begin{figure}[p!]
    \centering
    \includegraphics[width=0.95\textwidth]{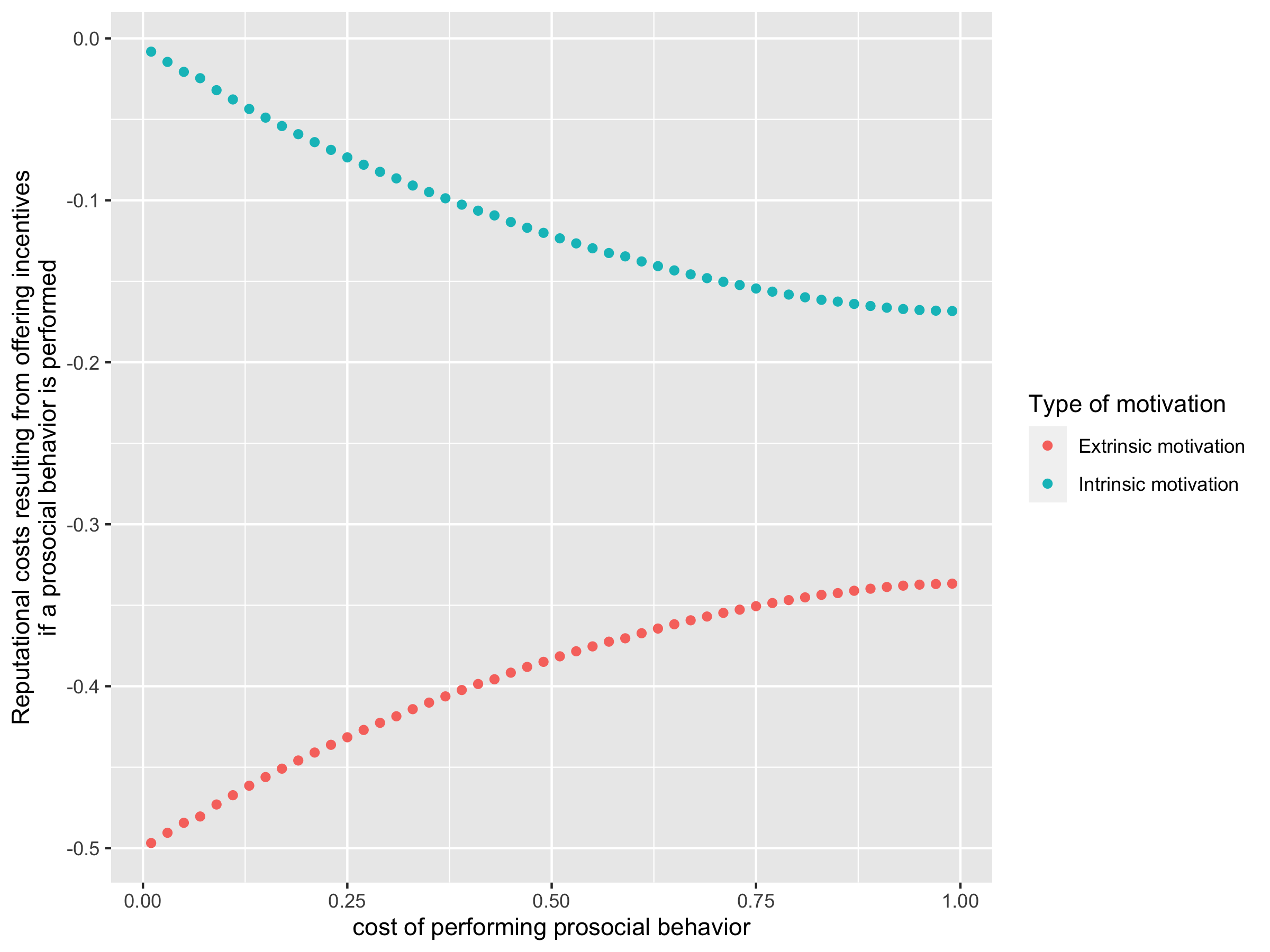}
    \caption{Line plot displaying the additional reputational \textit{cost} associated with performing a prosocial behavior in the presence of incentives, given BT's assumption that higher levels of expected extrinsic motivation represent a cost and higher levels of intrinsic motivation represent a benefit. That is, the blue line represents the benefit in terms of higher expected \textit{intrinsic} motivation when performing the behavior while receiving an incentive (i.e., $E(v_{a} | R = 1, B = 1) - E(v_{a} | R = 1, B = 0)$) \textbf{minus} the benefit in terms of higher expected intrinsic motivation when performing the behavior while \textit{not} receiving an incentive (i.e., $E(v_{a} | R = 0, B = 1) - E(v_{a} | R = 0, B = 0)$). In other words, the blue line captures the \textit{cost} of being perceived less intrinsically motivated when performing a prosocial behavior for which incentives are offered opposed to performing the behavior in the absence of incentives. The red line represents the cost in terms of higher expected extrinsic motivation when performing the behavior while receiving an incentive (i.e., $E(v_{v} | R = 1, B = 0) - E(v_{v} | R = 1, B = 1)$; note that it is not necessary to subtract the effect of expected extrinsic motivation when performing the behavior while \textit{not} receiving an incentive, because without incentives there is no information about an agent's extrinsic motivation, i.e. $E(v_{v} | R = 0, B = 0) - E(v_{v} | R = 0, B = 1) = 0$, see Eqn. \ref{eqn:SI3}).}
    \label{fig:model_rep}
\end{figure} \clearpage }

\afterpage{ \begin{figure}[p!]
    \centering
    \includegraphics[width=0.7\textwidth]{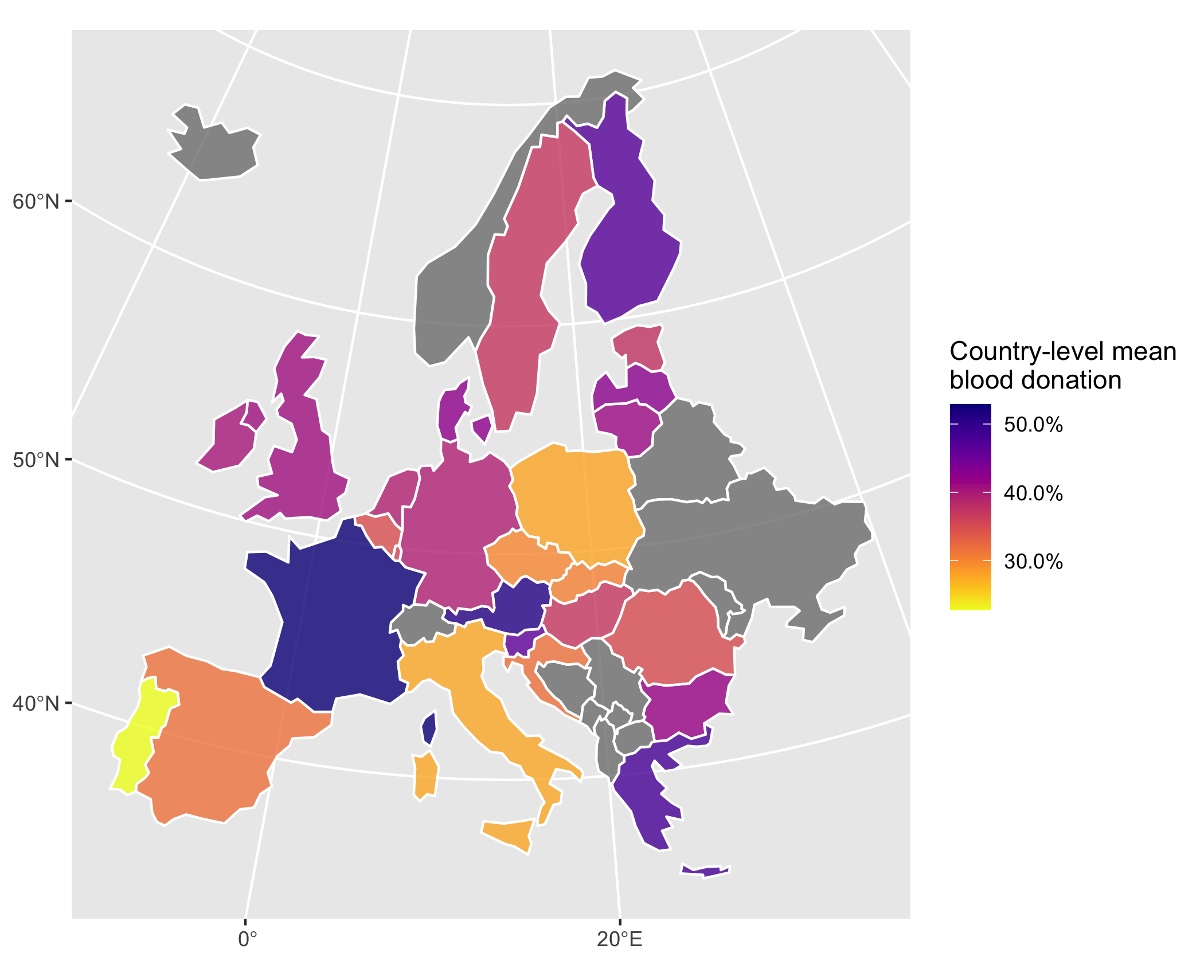}
    \caption{Country-level mean levels of blood donation.}
    \label{fig:map_BD}
\end{figure} \clearpage }

\afterpage{ \begin{figure}[p!]
\begin{subfigure}{.48\textwidth}
  \centering
  \includegraphics[width=\linewidth]{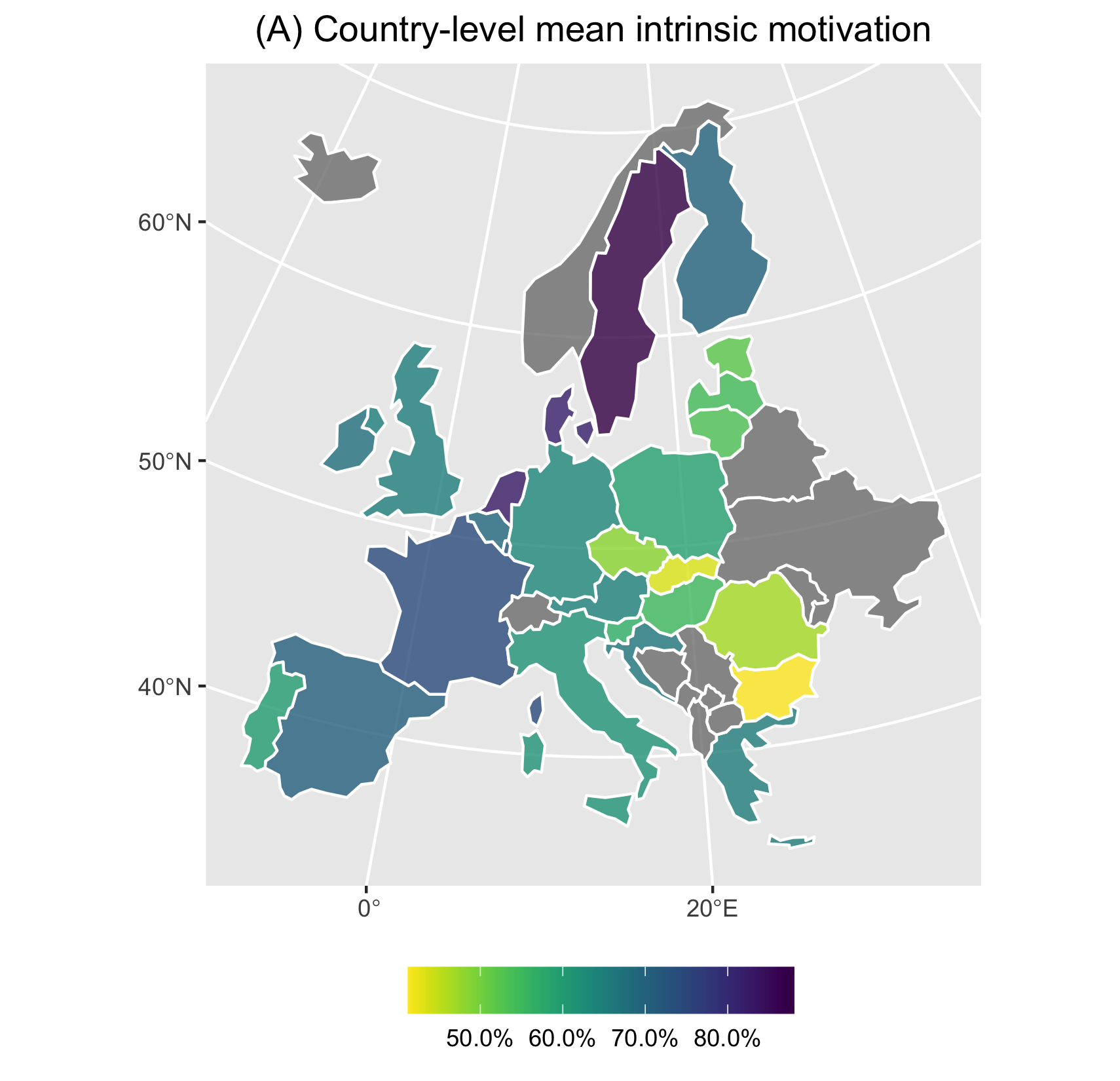}  
\end{subfigure}
\begin{subfigure}{.48\textwidth}
  \centering
  \includegraphics[width=\linewidth]{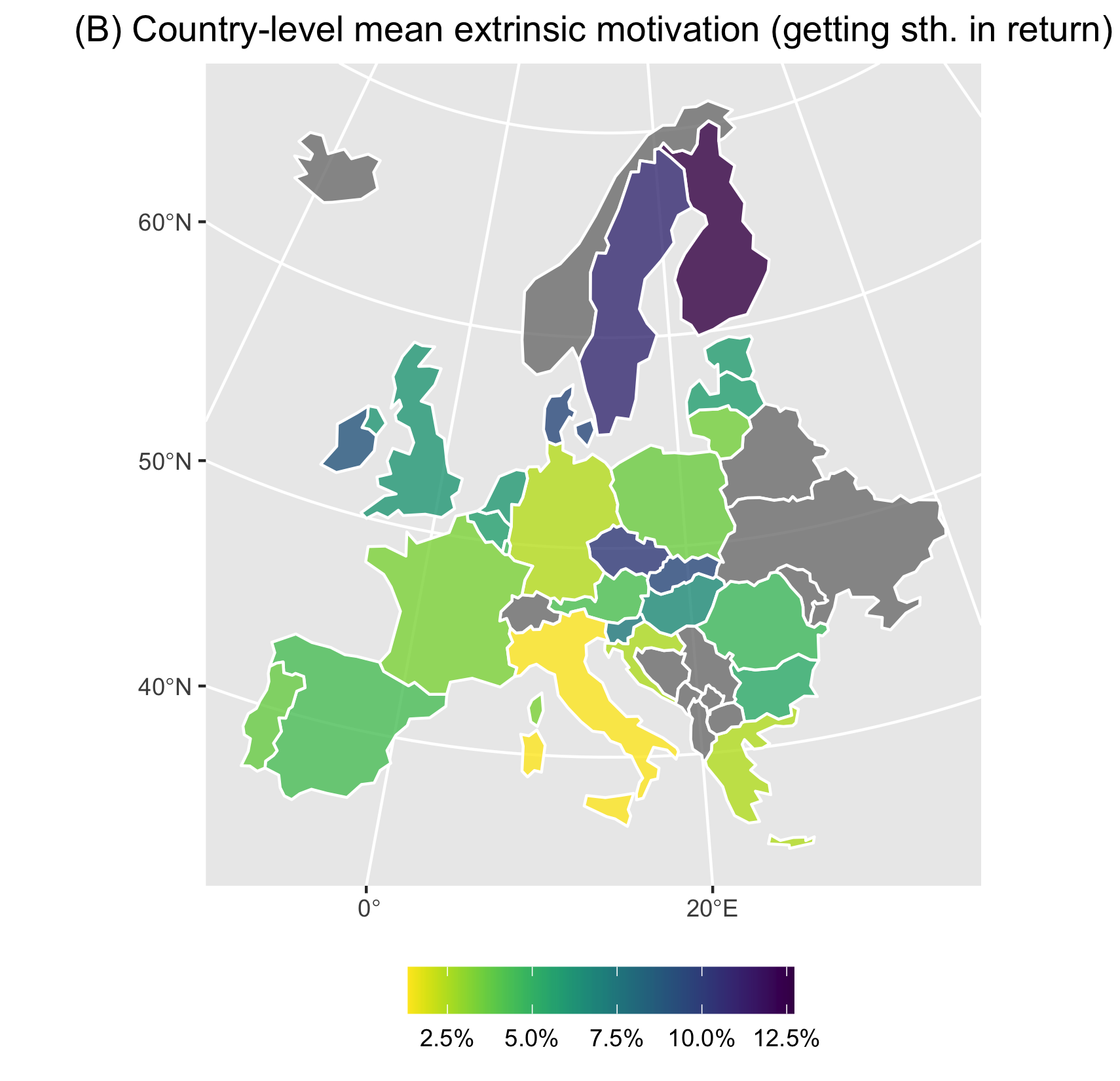}
\end{subfigure}
\caption{Country-level mean levels of (A) intrinsic motivation and (B) extrinsic motivation (in terms of mentioning “getting something in return” as a motivation to donate).}
\label{fig:map_IM_EM}
\end{figure} \clearpage }

\afterpage{ \begin{figure}[p!]
    \centering
    \includegraphics[width=0.9\textwidth]{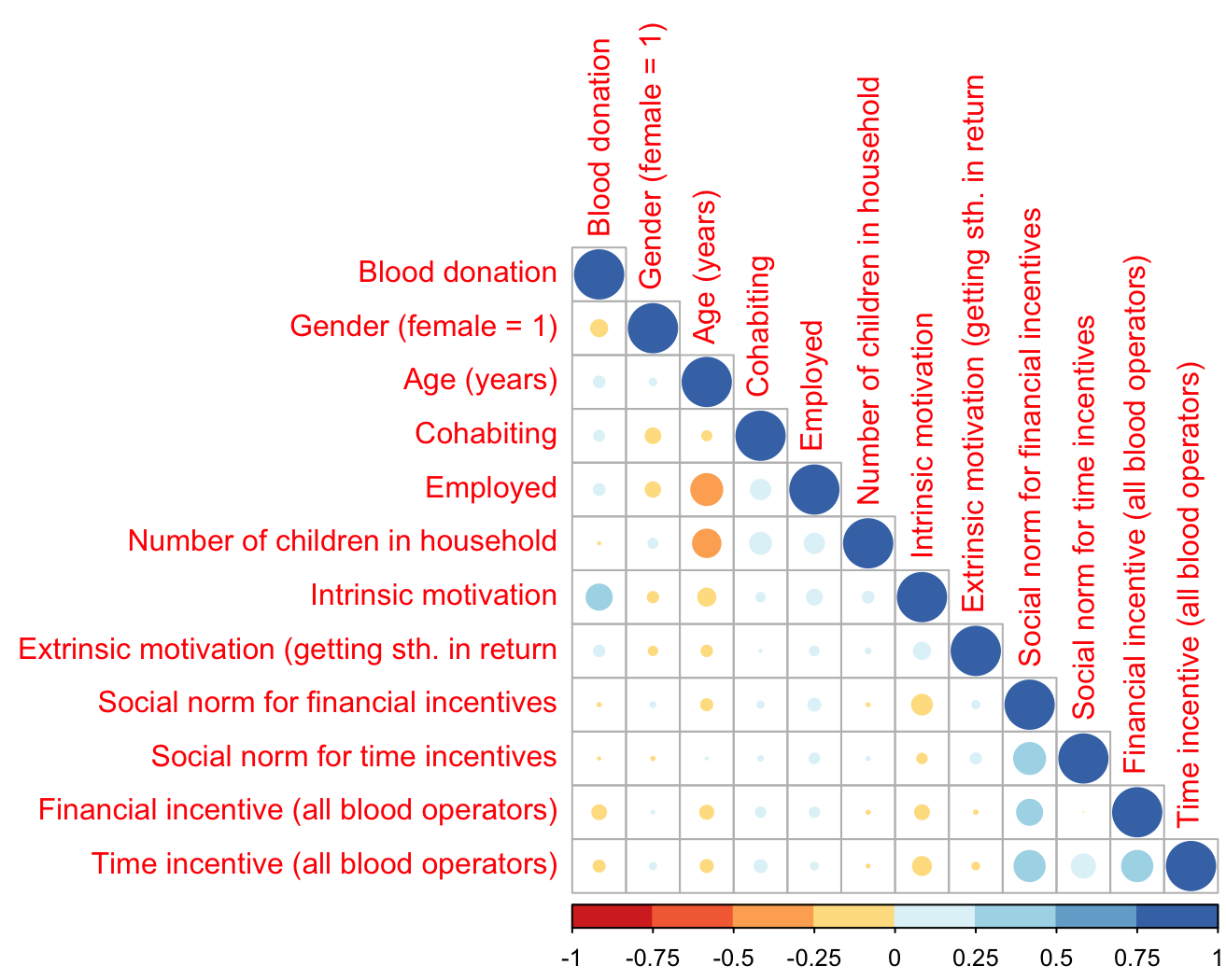}
    \caption{Correlation matrix of predictors of interest (only continuous and dichotomous variables). Correlation coefficients involving a dichotomous variable reflect the point-biserial coefficient; otherwise coefficients are Pearson's correlation coefficients.}
    \label{fig:corr_plot}
\end{figure} \clearpage }

\afterpage{ \begin{figure}[p!]
    \centering
    \includegraphics[width=0.9\textwidth]{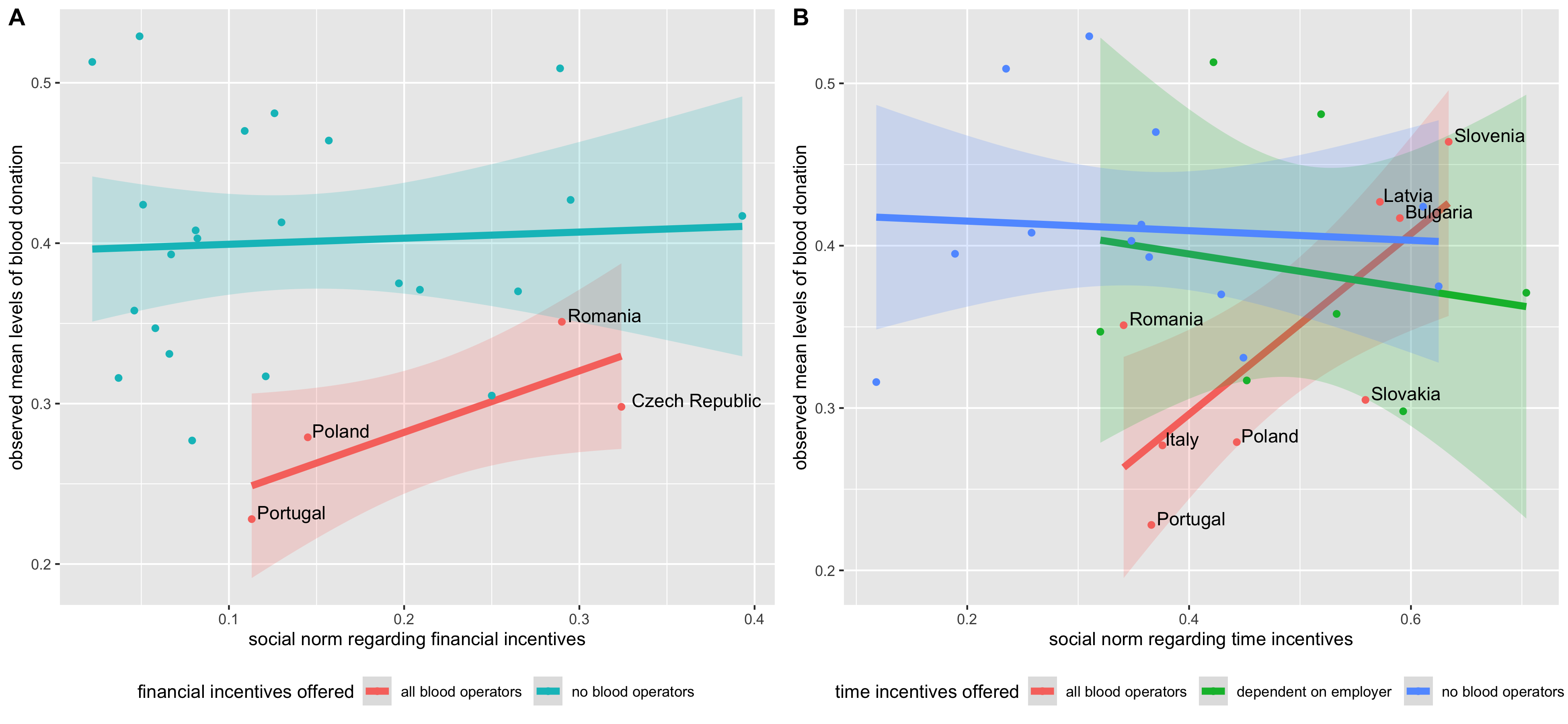}
    \caption{Country-level mean donation rates as a function of (A) social norms regarding \textit{financial} incentives (grouped by two levels of financial incentives) and (B) social norms regarding \textit{time} incentives (grouped by three levels of time incentives).}
    \label{fig:combined_descr}
\end{figure} \clearpage }

\afterpage{ \begin{figure}[p!]
\centering
\includegraphics[width=0.65\linewidth]{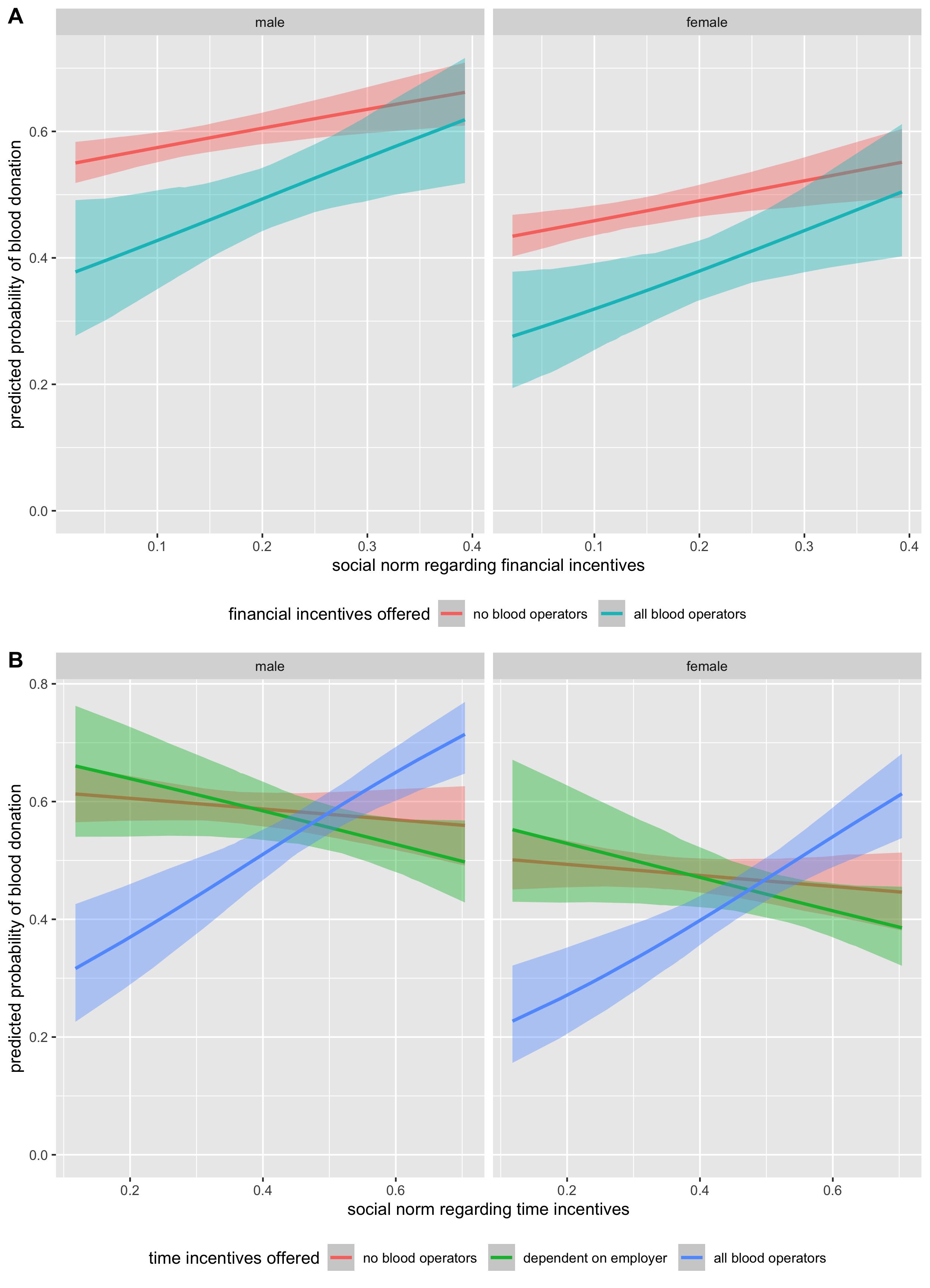}
\caption{Predictive margins of donating blood for a representative European man (left) and woman (right) (i.e., mean-aged, living with a partner, with secondary education, employed, living in a small town, with mean-number of children, rarely having difficulty paying bills, intrinsically but not extrinsically motivated) as a function of (A) social norms regarding \textit{financial} incentives (grouped by two levels of financial incentives) and (B) social norms regarding \textit{time} incentives (grouped by three levels of time incentives). Prediction bands indicate 80\% prediction intervals.}
\label{fig:pred_gender}
\end{figure} \clearpage }

\afterpage{ \begin{figure}[p!]
\centering
\includegraphics[width=0.65\linewidth]{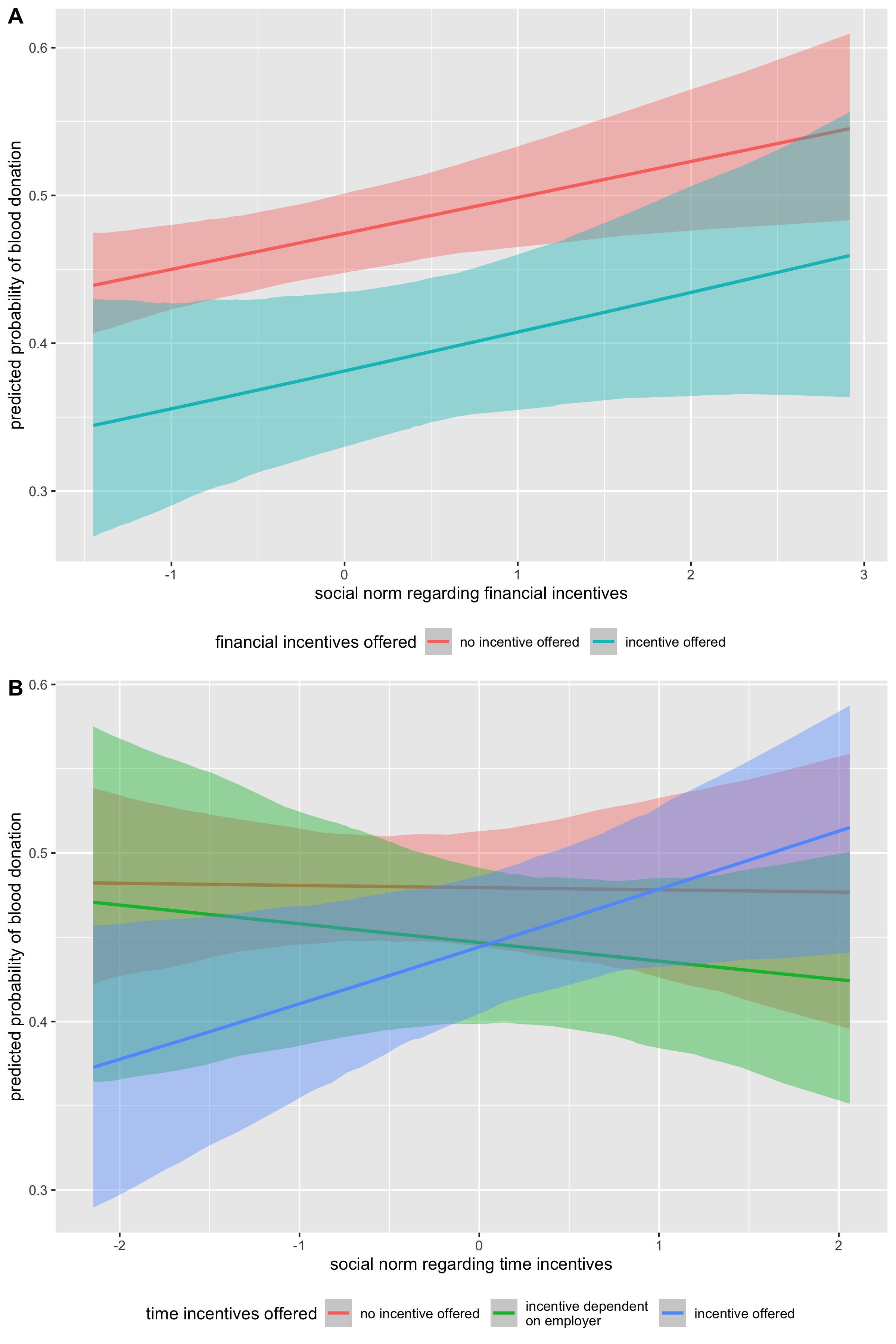}
\caption{Region-level exploratory analysis results: Predictive margins of donating blood for a representative European woman (i.e., mean-aged, living with a partner, with secondary education, employed, living in a small town, with mean-number of children, rarely having difficulty paying bills, intrinsically but not extrinsically motivated) as a function of \textbf{region-level} (A) social norms regarding \textit{financial} incentives (grouped by two levels of financial incentives at the country-level) and (B) social norms regarding \textit{time} incentives (grouped by three levels of time incentives at the country-level). Prediction bands indicate 80\% prediction intervals.}
\label{fig:region_scatter}
\end{figure} \clearpage }

\afterpage{ \begin{table}[p!]
\caption{Descriptive statistics for the dependent and independent variables.}
\centering
\begin{tabular}{l|l|l|l|l|l}
 & Variable & Range & Mean & Std. & n \\ \hline \hline
Individual-level & Blood donation & 0; 1 & 0.38 & - & 26532 \\ 
 & Gender (female = 1) & 0; 1 & 0.56 & - & 26532 \\ 
 & Age (years) & 18-99 & 51.29 & 17.81 & 26532 \\ 
 & Education & & & & \\ 
 & \hspace{0.5cm} None (ref.) & 0; 1 & 0.01 & - & 25640 \\ 
 & \hspace{0.5cm} $\leq 15$ years & 0; 1 & 0.17 & - & 25640 \\ 
 & \hspace{0.5cm} 16-19 years & 0; 1 & 0.45 & - & 25640 \\ 
 & \hspace{0.5cm} $\geq 20$ years & 0; 1 & 0.37 & - & 25640 \\ 
 & Cohabiting & 0; 1 & 0.65 & - & 26476 \\ 
 & Employed & 0; 1 & 0.49 & - & 26532 \\ 
 & Difficulty paying bills & & & & \\ 
 & \hspace{0.5cm} Never (ref.) & 0; 1 & 0.62 & - & 26192 \\ 
 & \hspace{0.5cm} Sometimes & 0; 1 & 0.27 & - & 26192 \\ 
 & \hspace{0.5cm} Most of the time & 0; 1 & 0.11 & - & 26192 \\ 
 & Intrinsic motivation & 0; 1 & 0.63 & - & 26532 \\ 
 & \begin{tabular}[c]{@{}l@{}}Extrinsic motivation \\ (getting sth. in return)\end{tabular} & 0; 1 & 0.06 & - & 26532 \\ 
 & \begin{tabular}[c]{@{}l@{}}Number of children in \\ household\end{tabular} & 0-15 & 0.29 & 0.7 & 26527 \\ 
 & Type of community & & & & \\ 
 & \hspace{0.5cm} Large town (ref.) & 0; 1 & 0.27 & - & 26518 \\ 
 & \hspace{0.5cm} Mid-sized town & 0; 1 & 0.42 & - & 26518 \\ 
 & \hspace{0.5cm} Rural area & 0; 1 & 0.31 & - & 26518 \\ \hline
Country-level & Financial incentives & & & & \\ 
 & \hspace{0.5cm} No blood operators (ref.) & 0; 1 & 0.82 & - & 28 \\ 
 & \hspace{0.5cm} Some blood operators & 0; 1 & 0.04 & - & 28 \\ 
 & \hspace{0.5cm} All blood operators & 0; 1 & 0.14 & - & 28 \\ 
 & Time incentives & & & & \\ 
 & \hspace{0.5cm} No blood operator (ref.) & 0; 1 & 0.46 & - & 28 \\ 
 & \hspace{0.5cm} Dependent on employer & 0; 1 & 0.25 & - & 28 \\ 
 & \hspace{0.5cm} Independent of employer & 0; 1 & 0.29 & - & 28 \\ 
 & \begin{tabular}[c]{@{}l@{}}Social norm \\ (financial incentives)\end{tabular} & 0.02-0.39 & 0.15 & 0.1 & 28 \\ 
 & \begin{tabular}[c]{@{}l@{}}Social norm \\ (time incentives)\end{tabular} & 0.12-0.70 & 0.43 & 0.15 & 28 \\ 
\end{tabular}
\label{table:descriptives}
\end{table} \clearpage }

\afterpage{ \begin{table}[p!]
\centering
\caption{Description of financial and time incentives offered to blood donors in 28 European countries. 0 denotes \textit{offered by no blood operators}, 0.5 denotes \textit{offered only by some blood operators or only to some donors} and 1 denotes \textit{offered by all blood operators}.}
\begin{tabularx}{\textwidth}{c|c|Y|c|Y}
country & Financial incentive & Description & Time incentive & Description \\ \hline \hline
Austria & 0 & & 0 & \\ \hline
\multirow{2}{*}{Belgium} & \multirow{2}{*}{0} &  & \multirow{2}{*}{0.5} & depends on employers;\linebreak up to 1 day off \\ \hline
\multirow{2}{*}{Bulgaria} & \multirow{2}{*}{0} & & \multirow{2}{*}{1} & not depending on employer; 2 days \\ \hline
\multirow{2}{*}{Croatia} & \multirow{2}{*}{0} & & \multirow{2}{*}{0.5} & depends on employers; \linebreak time of donation \\ \hline
\multirow{2}{*}{Cyprus} & \multirow{2}{*}{0} & & \multirow{2}{*}{0.5} & depends on employers;\linebreak time of donation \\ \hline
\multirow{3}{*}{Czech Republic} & \multirow{3}{*}{1} & \multirow{3}{*}{tax relief 17–18 Euro} & \multirow{3}{*}{0.5} & depends on employers; duration of donation + travel \\ \hline
Denmark & 0 & & 0 & \\ \hline
Estonia & 0 & & 0 & \\ \hline
Finland & 0 & & 0 & \\ \hline
France & 0 & & 0 & \\ \hline
\multirow{3}{*}{Germany} & \multirow{3}{*}{0.5} & 25 Euro cash (only some blood operators, not Red Cross) & \multirow{3}{*}{0} & \\ \hline
\multirow{3}{*}{Greece} & \multirow{3}{*}{0} & & \multirow{3}{*}{0.5} & depends on employers, especially civil servants; 2 days \\ \hline
Hungary & 0 & & 0 & \\ \hline
Ireland & 0 & & 0 & \\ \hline
\multirow{2}{*}{Italy} & \multirow{2}{*}{0} & & \multirow{2}{*}{1} & not dependent on employer \\ \hline
\multirow{2}{*}{Latvia} & \multirow{2}{*}{0} &  & \multirow{2}{*}{1} & not dependent on employer; 1 day \\ \hline
Lithuania & 0 & & 0 & \\ \hline
\multirow{4}{*}{Luxembourg} & \multirow{4}{*}{0} & & \multirow{4}{*}{0.5} & depends on employers; especially for public sector; duration of donation: $\sim$1 hour \\ \hline
Malta & 0 & & 0 & \\ \hline
Netherlands & 0 & & 0 & \\ \hline
\multirow{2}{*}{Poland} & \multirow{2}{*}{1} & \multirow{2}{*}{tax relief; 29,40 Euro/L} & \multirow{2}{*}{1} & not dependent on employer; 1 day \\ \hline
\multirow{4}{*}{Portugal} & \multirow{4}{*}{1} & exemption from paying fees for access to the national healthcare (tax cut) & \multirow{4}{*}{1} & \vspace{0.1cm} duration of donation: $\sim$1 hour \\ \hline
\multirow{2}{*}{Romania} & \multirow{2}{*}{1} & \multirow{2}{*}{food vouchers; \linebreak 13 Euro} & \multirow{2}{*}{1} & not dependent on employer; 1 day \\ \hline
\multirow{2}{*}{Slovakia} & \multirow{2}{*}{0} & & \multirow{2}{*}{1} & not dependent on employer; up to 1 day \\ \hline
Slovenia & 0 & & 1 & 1 day off \\ \hline
Spain & 0 & & 0 & \\ \hline
Sweden & 0 & & 0.5 & depends on employers \\ \hline
United Kingdom & 0 & & 0 & \\ \end{tabularx}
\label{tab:incentives}
\end{table} \clearpage }

\afterpage{ \begin{table}[p!]
\centering
\caption{Model results for financial incentives (full sample and robustness check excluding respondents who indicated they are not willing to donate). Standard errors in brackets; *** denotes p $<$ 0.001; ** denotes p $<$ 0.01; * denotes p $<$ 0.05; $\diamond$ indicates variable was normalized.}
\begin{tabular}{l|p{7.5cm}|l|l}
 & Variable & (1) Full sample & (2) Rob. check \\ \hline \hline
Individual-level & (Intercept) & -1.302 (0.17) *** & 0.029 (0.182)\\ 
 & Age $\diamond$  & 0.312 (0.019) *** & 0.394 (0.02) ***\\ 
 & Gender (Female = 1) & -0.465 (0.028) *** & -0.457 (0.03) ***\\ 
 & Cohabiting & 0.079 (0.031) * & 0.038 (0.033)\\ 
 & Education &  & \\ 
 & \hspace{0.7cm} No full-time education (ref.) & - & -\\ 
 & \hspace{0.7cm} $\leq 15$ years & 0.06 (0.154) & 0.094 (0.165)\\ 
 & \hspace{0.7cm} 16-19 years & 0.289 (0.152) & 0.234 (0.163)\\ 
 & \hspace{0.7cm} $\geq 20$ years & 0.51 (0.152) *** & 0.453 (0.163) **\\ 
 & Not employed & -0.238 (0.033) *** & -0.209 (0.034) ***\\ 
 & Difficulty paying bills &  & \\ 
 & \hspace{0.7cm} Never (ref.) & - & -\\ 
 & \hspace{0.7cm} Sometimes & 0.027 (0.035) & 0.041 (0.037)\\ 
 & \hspace{0.7cm} Most of the time & 0.041 (0.051) & 0.094 (0.055)\\ 
 & Number of children in household $\diamond$ & 0.019 (0.015) & 0.02 (0.015)\\ 
 & Type of community & & \\ 
 & \hspace{0.7cm} Large town (ref.) & - & -\\ 
 & \hspace{0.7cm} Mid-sized town & -0.046 (0.035) & -0.051 (0.037)\\ 
 & \hspace{0.7cm} Rural area & -0.091 (0.038) * & -0.099 (0.041) *\\ 
 & Intrinsic motivation & 1.346 (0.033) *** & 0.091 (0.038) *\\ 
 & Extrinsic motivation (getting sth. in return) & 0.11 (0.062) & -0.067 (0.062)\\ \hline 
Country-level & Financial incentive &  & \\ 
 & \hspace{0.7cm} No blood operator (0; ref.) & - & -\\ 
 & \hspace{0.7cm} All blood operators (1) & -0.512 (0.198) ** & -0.485 (0.204) *\\ 
 & Social norm for fin. incentives $\diamond$ & 0.125 (0.063) * & 0.174 (0.065) **\\ 
 & Social norm for fin. incentives $\diamond$ $\times$ fin. incentive &  & \\ 
 & \hspace{0.7cm} Social norm $\times$ incentive = 0 (ref.) & - & -\\ 
 & \hspace{0.7cm} Social norm $\times$ incentive = 1 & 0.137 (0.178) & 0.099 (0.183)\\ \hline
Cross-level & Extrinsic motivation $\times$ fin. incentive &  & \\ 
interaction & \hspace{0.7cm} Extrinsic motivation $\times$ incentive = 0 (ref.) & - & -\\ 
 & \hspace{0.7cm} Extrinsic motivation $\times$ incentive = 1 & 0.283 (0.168) & 0.183 (0.167)\\ \hline
 & N & 23910 & 18961
\end{tabular}
\label{tab:model_financial}
\end{table} \clearpage }

\afterpage{ \begin{table}[p!]
\centering
\caption{Model results for time incentives (full sample and robustness check excluding respondents who indicated they are not willing to donate). Standard errors in brackets; *** denotes p $<$ 0.001; ** denotes p $<$ 0.01; * denotes p $<$ 0.05; $\diamond$ indicates variable was normalized.}
\begin{tabular}{l|p{7.3cm}|l|l}
 & Variable & (1) Full sample & (2) Rob. check\\ \hline \hline
Individual-level & (Intercept) & -1.564 (0.177) *** & -0.168 (0.192) \\
 & Age $\diamond$ & 0.316 (0.018) *** & 0.394 (0.019) *** \\
 & Gender (Female = 1) & -0.455 (0.027) *** & -0.445 (0.029) *** \\
 & Cohabiting & 0.076 (0.03) * & 0.038 (0.032) \\
 & Education &  & \\
 & \hspace{0.7cm} No full-time education (ref.) & - & -\\
 & \hspace{0.7cm} $\leq 15$ years & 0.053 (0.151) & 0.090 (0.163) \\
 & \hspace{0.7cm} 16-19 years & 0.258 (0.149) & 0.200 (0.161) \\
 & \hspace{0.7cm} $\geq 20$ years & 0.483 (0.149) ** & 0.419 (0.161) ** \\
 & Employed & 0.244 (0.031) *** & 0.213 (0.033) *** \\
 & Number of children in household $\diamond$ & 0.021 (0.014) & 0.022 (0.015) \\
 & Type of community &  & \\
 & \hspace{0.7cm} Large town (ref.) & - & -\\
 & \hspace{0.7cm} Mid-sized town & -0.039 (0.034) & -0.044 (0.036) \\
 & \hspace{0.7cm} Rural area & -0.068 (0.037) & -0.075 (0.039) \\
 & Intrinsic motivation & 1.369 (0.032) *** & 0.093 (0.037) * \\
 & Extrinsic motivation (getting sth. in return) & 0.103 (0.079) & -0.033 (0.078) \\ \hline
Country-level & Time incentive &  & \\
 & \hspace{0.7cm} No blood operator (0; ref.) & - & -\\
 & \hspace{0.7cm} Dependent on employer (0.5) & -0.033 (0.165) & -0.101 (0.179) \\
 & \hspace{0.7cm} Independent of employer (1) & -0.236 (0.151) & -0.201 (0.163) \\
 & Social norm for time incentives $\diamond$ & -0.055 (0.086) & -0.054 (0.093) \\
 & Social norm for time incentives $\diamond$ $\times$ incentive &  & \\
 & \hspace{0.7cm} Social norm $\times$ incentive = 0 (ref.) & - & -\\
 & \hspace{0.7cm} Social norm $\times$ incentive = 0.5 & -0.116 (0.169) & -0.113 (0.183) \\
 & \hspace{0.7cm} Social norm $\times$ incentive = 1 & 0.483 (0.168) ** & 0.479 (0.182) ** \\ \hline
Cross-level & Extrinsic motivation $\times$ time incentive &  & \\
interaction & \hspace{0.7cm} Extrinsic motivation $\times$ incentive = 0 (ref.) & - & -\\
 & \hspace{0.7cm} Extrinsic motivation $\times$ incentive = 0.5 & 0.068 (0.144) & 0.092 (0.142) \\
 & \hspace{0.7cm} Extrinsic motivation $\times$ incentive = 1 & 0.107 (0.139) & -0.111 (0.137) \\ \hline
 & N & 25586 & 20218
\end{tabular}
\label{tab:model_time}
\end{table} \clearpage }

\newpage

\begin{footnotesize}
\begin{longtable}[p!]{p{2cm}|p{2.5cm}|p{4.5cm}|p{4.5cm}}
\caption{Overview of dependent and independent variables. $\triangleright$ denotes source: Eurobarometer; $\star$ denotes source: Expert survey.} \\
Variable & Description & Original item & Response range \\ \hline \hline
Blood donation &Individual-level response: Blood donation during lifetime $\triangleright$  & During the lifetime of a person it is possible to donate different body substances (blood or cells) to help other people. Could you please indicate which ones you have [...] donate[d] yourself? \newline \newline Blood & $0; 1$ (dummy coded) \newline \newline 0 (not donated blood in the past) \newline \newline 1 (donated blood in the past) \\ \hline
Intrinsic motivation & Donation motivated by intrinsic factors (i.e., help people in need, alleviate shortages, medical research) $\triangleright$  & You said that you have or would be prepared to donate certain body substances during your lifetime (blood or cells) or after death (tissues). For which of the following reasons have you or would you donate any of the body substances mentioned earlier?\newline \newline  To help other people in need\newline \newline To alleviate shortages of these substances\newline \newline  To support medical research & $0; 1$ (dummy coded)\newline \newline  0 (did not mention intrinsic motivator)\newline \newline  1 (mentioned intrinsic motivator) \\ \hline
Extrinsic motivation & Donation motivated by extrinsic factors (i.e., to receive something in return) $\triangleright$  & You said that you have or would be prepared to donate certain body substances during your lifetime (blood or cells) or after death (tissues). For which of the following reasons have you or would you donate any of the body substances mentioned earlier?\newline \newline  To receive something in return for you or your relatives & $0; 1$ (dummy coded)\newline \newline  0 (did not mention extrinsic motivator)\newline \newline  1 (mentioned extrinsic motivator) \\ \cline{2-4}
 &Subjective difficulty to pay bills $\triangleright$  & During the last twelve months, would you say you had difficulties to pay your bills at the end of the month...? & $0; 1; 2$ (dummy coded)\newline \newline 0 (Never/Almost never)\newline \newline  1 (From time to time)\newline \newline  2 (Most of the time) \\ \cline{2-4} 
 &Lack of current occupation $\triangleright$  & What is your current occupation? & $0; 1$ (dummy coded)\newline \newline  0 (other)\newline \newline  1 (Homemaker, student, unemployed, retired or unable to work through illness) \\ \hline
Personal cost &  Time constraints due to child care $\triangleright$  & Could you tell me how many children less than 10 years old live in your household? & numeric \\ \cline{2-4} 
 &Travel costs due to rural living $\triangleright$  & Would you say you live in a...\newline \newline  Rural area or village\newline \newline  Small or middle sized town\newline \newline  Large town & $0; 1; 2$ (dummy coded)\newline \newline  0 (lives in large town)\newline \newline  1 (lives in small or middle sized town)\newline \newline  2 (lives in rural area or village) \\ \hline
Social norms &  Aggregated response at country-level: \newline Mean social norm regarding financial incentives for blood donation $\triangleright$  & For donating blood or plasma during someone’s lifetime, do you consider it acceptable to receive cash amounts additional to the reimbursement of the costs related to the donation? & $[0, 1]$ (aggregated from individual-level $i \in \{0; 1\}$)\newline \newline  0 (0\% of population found acceptable)\newline \newline  1 (100\% of population found acceptable) \\ \cline{2-4} 
 &Aggregated response at country-level: \newline Mean social norm regarding time incentives for blood donation $\triangleright$  & For donating blood or plasma during someone’s lifetime, do you consider it acceptable to receive time off work (for the time needed for the donation and/or for recovery) & $[0, 1]$ (aggregated from individual-level $i \in \{0; 1\}$)\newline \newline  0 (0\% of population found acceptable)\newline \newline  1 (100\% of population found acceptable) \\ \hline
Controls & Age of respondent $\triangleright$  & How old are you? & numeric \\ \cline{2-4} 
 &Gender $\triangleright$  & NA & $0; 1$ (dummy coded)\newline \newline 0 (Male)\newline \newline  1 (Female) \\ \cline{2-4} 
 &Education (Age when finished full-time education) $\triangleright$  & How old were you when you stopped full-time education? & $0; 1; 2; 3$ (dummy coded)\newline \newline  0 (No full-time education)\newline \newline  1 (15 years or younger)\newline \newline  2 (16-19 years)\newline \newline  3 (20 years and older) \\ \cline{2-4} 
 &Partner status (live with partner or not) $\triangleright$  & Could you give me the letter which corresponds best to your own current situation?\newline \newline  (recoded from 14 levels, e.g., married/remarried and living with/without children) & $0; 1$ (dummy coded)\newline \newline  0 (Living without partner)\newline \newline  1 (Living with partner) \\ \hline
Incentives & Country-level variable for whether financial incentives are offered for donating blood. $\star$  & See expert survey & $0; 0.5; 1$ (dummy coded)\newline \newline  0 (no blood establishments offer financial incentives for donating blood)\newline \newline  0.5 (some blood establishments offer financial incentives for donating blood)\newline \newline  1 (all blood establishments offer financial incentives for donating blood) \\ \cline{2-4} 
 &Country-level variable for whether time incentives are offered for donating blood. $\star$  & See expert survey & $0; 0.5; 1$ (dummy coded)\newline \newline  0 (no time incentives are offered for donating blood)\newline \newline  0.5 (time incentives are offered for donating blood dependent on employer)\newline \newline  1 (time incentives are offered for donating blood independent of employer)\\ 
 \label{tab:vars}
 \end{longtable}
\end{footnotesize}



\section*{Supplementary references}

\begin{enumerate}
  \setcounter{enumi}{53}
  \item A. Diekmann and P. Preisendörfer, “Green and greenback: The behavioral effects of environmental attitudes in low-cost and high-cost situations,” \textit{Rationality and Society}, vol. 15, no. 4, pp. 441–472, 2003. Publisher: Sage Publications
  \item C. Weidmann and H. Klüter, “Blood collection and donor motivation in Germany,” \textit{ISBT Science Series}, vol. 8, no. 1, pp. 238–241, 2013. Publisher: Wiley Online Library.
\end{enumerate}

\end{document}